\documentclass[pra,aps,preprint,superscriptaddress,showpacs,nofootinbib]{revtex4-1}

\usepackage[dvips]{graphicx}
\usepackage{amsmath,amssymb,amsthm,mathrsfs,amsfonts,dsfont}
\usepackage{subfigure, epsfig}
\usepackage{braket}
\usepackage{bm}
\usepackage{enumerate}

\newtheorem{definition}{Definition}

\newcommand{\ea}{\mathrm{EA}}
\newcommand{\oexp}{\omega_\mathrm{exp}}
\newcommand{\dest}{\delta_\mathrm{est}}

\usepackage{graphicx}   
\usepackage{color}
\usepackage[10pt]{moresize}

\usepackage{amscd}

\usepackage{epsfig}
\usepackage{subfigure}
\usepackage{graphicx}

\begin{document}

\title{High speed self-testing quantum random number generation without detection loophole}

\author{Yang Liu}
\affiliation{Shanghai Branch, National Laboratory for Physical Sciences at Microscale and Department of Modern Physics, University of Science and Technology of China, Shanghai 201315, P.~R.~China}
\affiliation{Shanghai Branch, CAS Center for Excellence and Synergetic Innovation Center in Quantum Information and Quantum Physics, University of Science and Technology of China, Shanghai 201315, P.~R.~China}

\author{Xiao Yuan}
\affiliation{Shanghai Branch, National Laboratory for Physical Sciences at Microscale and Department of Modern Physics, University of Science and Technology of China, Shanghai 201315, P.~R.~China}
\affiliation{Shanghai Branch, CAS Center for Excellence and Synergetic Innovation Center in Quantum Information and Quantum Physics, University of Science and Technology of China, Shanghai 201315, P.~R.~China}
\affiliation{Center for Quantum Information, Institute for Interdisciplinary Information Sciences, Tsinghua University, Beijing 100084, P.~R.~China}

\author{Ming-Han Li}
\affiliation{Shanghai Branch, National Laboratory for Physical Sciences at Microscale and Department of Modern Physics, University of Science and Technology of China, Shanghai 201315, P.~R.~China}
\affiliation{Shanghai Branch, CAS Center for Excellence and Synergetic Innovation Center in Quantum Information and Quantum Physics, University of Science and Technology of China, Shanghai 201315, P.~R.~China}

\author{Weijun Zhang}
\affiliation{State Key Laboratory of Functional Materials for Informatics, Shanghai Institute of Microsystem and Information Technology, Chinese Academy of Sciences, Shanghai 200050, P.~R.~China}

\author{Qi Zhao}
\affiliation{Center for Quantum Information, Institute for Interdisciplinary Information Sciences, Tsinghua University, Beijing 100084, P.~R.~China}

\author{Jiaqiang Zhong}
\affiliation{Purple Mountain Observatory and Key Laboratory of Radio Astronomy, Chinese Academy of Sciences, 2 West Beijing Road, Nanjing, Jiangsu 210008, P.~R.~China}

\author{Yuan Cao}
\author{Yu-Huai Li}
\author{Luo-Kan Chen}

\affiliation{Shanghai Branch, National Laboratory for Physical Sciences at Microscale and Department of Modern Physics, University of Science and Technology of China, Shanghai 201315, P.~R.~China}

\affiliation{Shanghai Branch, CAS Center for Excellence and Synergetic Innovation Center in Quantum Information and Quantum Physics, University of Science and Technology of China, Shanghai 201315, P.~R.~China}

\author{Hao Li}
\affiliation{State Key Laboratory of Functional Materials for Informatics, Shanghai Institute of Microsystem and Information Technology, Chinese Academy of Sciences, Shanghai 200050, P.~R.~China}

\author{Tianyi Peng}
\affiliation{Center for Quantum Information, Institute for Interdisciplinary Information Sciences, Tsinghua University, Beijing 100084, P.~R.~China}

\author{Yu-Ao Chen}
\author{Cheng-Zhi Peng}
\affiliation{Shanghai Branch, National Laboratory for Physical Sciences at Microscale and Department of Modern Physics, University of Science and Technology of China, Shanghai 201315, P.~R.~China}
\affiliation{Shanghai Branch, CAS Center for Excellence and Synergetic Innovation Center in Quantum Information and Quantum Physics, University of Science and Technology of China, Shanghai 201315, P.~R.~China}

\author{Sheng-Cai Shi}
\affiliation{Purple Mountain Observatory and Key Laboratory of Radio Astronomy, Chinese Academy of Sciences, 2 West Beijing Road, Nanjing, Jiangsu 210008, P.~R.~China}

\author{Zhen Wang}
\author{Lixing You}
\affiliation{State Key Laboratory of Functional Materials for Informatics, Shanghai Institute of Microsystem and Information Technology, Chinese Academy of Sciences, Shanghai 200050, P.~R.~China}

\author{Xiongfeng Ma}
\affiliation{Center for Quantum Information, Institute for Interdisciplinary Information Sciences, Tsinghua University, Beijing 100084, P.~R.~China}

\author{Jingyun Fan}
\author{Qiang Zhang}
\author{Jian-Wei Pan}
\affiliation{Shanghai Branch, National Laboratory for Physical Sciences at Microscale and Department of Modern Physics, University of Science and Technology of China, Shanghai 201315, P.~R.~China}
\affiliation{Shanghai Branch, CAS Center for Excellence and Synergetic Innovation Center in Quantum Information and Quantum Physics, University of Science and Technology of China, Shanghai 201315, P.~R.~China}

\begin{abstract}
Quantum mechanics provides means of generating genuine randomness that is impossible with deterministic classical processes. Remarkably, the unpredictability of randomness can be certified in a self-testing manner that is independent of implementation devices. Here, we present an experimental demonstration of self-testing quantum random number generation based on an detection-loophole free Bell test with entangled photons. In the randomness analysis, without the assumption of independent identical distribution, we consider the worst case scenario that the adversary launches the most powerful attacks against quantum adversary. After considering statistical fluctuations and applying an 80 Gb $\times$ 45.6 Mb Toeplitz matrix hashing, we achieve a final random bit rate of 114 bits/s, with a failure probability less than $10^{-5}$. Such self-testing random number generators mark a critical step towards realistic applications in  cryptography and fundamental physics tests.
\end{abstract}

\maketitle

{\it Introduction.---}
Random numbers are widely used in applications ranging from numerical simulation and cryptography to lottery. While the foremost property of random number generators (RNGs) in many applications is distribution uniformity of its outputs, secure information processing applications such as cryptography demand additionally that the devices to produce randomness must be secure against any adversaries, regardless of classical or quantum mechanics. Classical RNGs have a deterministic nature and hence are not random. Quantum random number generators (QRNGs) rely on the unpredictability in breaking quantum coherence and are theoretically unpredictable. However, the unpredictability may be jeopardized in practice because the adversary may gain information about the devices and even maliciously manipulate the devices of QRNGs, which is often undetected by a finite set of statistical tests. Self-testing QRNGs certify the randomness unconditionally based on the loophole free violation of Bell's inequality, offering a reliable way of generating genuine randomness in a device-independent manner and therefore holding great promise for future applications. (See Ref.~\cite{ma2016quantum, RevModPhys.89.015004} for a review of QRNGs.)

Considering a loophole free Bell test experiment. Alice and Bob are honest parties at two remote sites, each receives one of a pair of entangled photons and measures its quantum state with a randomly selected measurement setting. Alice and Bob may not trust the devices because the devices may be prepared by the adversary. The experiment observes no-signaling theorem, $i.e.$, even at the speed of light, no information of measurement setting is conveyed to the source prior to the emission of entangled photon pairs (to close randomness loophole) and no information about Alice's (Bob's) measurement setting and measurement outcomes are conveyed to Bob (Alice) prior to his (her) state measurement (to close locality loophole). The photon-detection efficiency is sufficiently high for the experiment to be free from the detection loophole. Local hidden variable theories (based on classical determinism) set a bound to the correlation measurement between Alice and Bob. Breaking the bound exhibits quantum correlation, which cannot be explained by classical deterministic mechanisms. This nonlocal quantum correlation certifies that Alice and Bob's measurement outcomes possess genuine quantum randomness which is unaccessible to the adversary, irrelevant to the implementation devices \cite{Colbeck09, Colbeck12}.

As we deepen the understanding of self-testing QRNG \cite{Fehr13,Pironio13, Vazirani14, Miller14, Chung14, coudron2014infinite, Miller15, dupuis2016entropy, arnon2016simple}, the security analysis becomes more efficient in producing randomness and more robust to noise. In particular, the entropy accumulation theorem formulated by Dupuis, Fawzi, and Renner \cite{dupuis2016entropy} converts a single-short result to the multiple-shot case. Exploiting the entropy accumulation theorem, Arnon-Friedman, Renner and Vidick proposed a self-testing QRNG analysis method without the assumption of independent identical distribution (i.i.d.), which nevertheless produces randomness with yield approaching the value for the i.i.d. case \cite{arnon2016simple}. The analysis is also against quantum adversary. So far there were two reported experimental studies on self-testing QRNG. One was based on a detection loophole free Bell test experiment with entangled ions \cite{pironio2010},  which produced random bits at a rate of $1.5 \times 10^{-5}$ bit $s^{-1}$ without the assumption of i.i.d. The analysis is against classical adversary. The other one was based on a detection loophole free Bell test experiment with entangled photons \cite{Christensen13}, which produced random bits at a rate of 0.4 bit $s^{-1}$, albeit with the assumption of i.i.d.

Very recently, several experiments demonstrated the violation of Bell's inequality with both locality and detection loopholes closed simultaneously \cite{hensen2015loophole, Shalm15, Giustina15, Rosenfeld16}. It was also shown that the randomness loophole can be progressively addressed with cosmic RNGs \cite{Gallicchio14,Handsteiner17,cheng2016random}, which take advantage of randomness at the remote celestial object to set the time constraint of local hidden variable mechanisms deep into the cosmic history. These works pave the way to construct a practical self-testing QRNG.  Here we report an experimental realization of a self-testing QRNG based on a detection loophole free Bell test with entangled photon pairs, with randomness extraction of 114 bits $s^{-1}$ and uniformity within $10^{-5}$, which is a significant improvement over the previous experiment \cite{pironio2010}. 
The randomness generation analysis is against the most powerful quantum adversary attacks and does not rely on the independent identical distribution assumption. Our experiment marks a critical step for generating self-testing QRNGs for practical applications.

{\it Proposal.---}
The self-testing QRNG protocol is based on the Bell test experiment, namely a Clauser-Horne-Shimony-Holt game \cite{CHSH}. In each experimental trial, Alice and Bob perform state measurements upon receiving random inputs $x$ and $y$ and produce outputs $a$ and $b$, respectively. According to local hidden variable models, the correlations described by probability distributions $p(ab|xy)$ in the i.i.d. scenario are factorable with $p(ab|xy)=\sum_\lambda p(\lambda)p(a|x,\lambda)p(b|y,\lambda)$. The $J$-value of the CHSH game satisfies an inequality,
\begin{equation}\label{Eq:CHSHvalue}
	J = \frac{1}{4}\sum_{abxy}\beta_{abxy}p(ab|xy)-3/4\le 0,
\end{equation}
where the pay-off coefficient is given by
\begin{equation}\label{Eq:payoff}
	\beta_{abxy} = \begin{cases} 1, & \mbox{if } a\oplus b=x\cdot y \\ 0 & \mbox{if } a\oplus b\neq x\cdot y \end{cases},
\end{equation}
with $\oplus$ standing for plus modulo 2.  Quantum theory allows $J > 0$ as opposed to local hidden variable models.

In practice, all Bell test experiments have finite statistics. Instead of approximating probability distributions based on the i.i.d. assumption, we introduce the Bell value $J_i$ in experimental trial $i$,
\begin{equation}\label{Eq:Jpayoff}
	J_i = \begin{cases} 1, & \mbox{if } a_i\oplus b_i=x_i\cdot y_i \\ 0, & \mbox{ } otherwise \end{cases}.
\end{equation}
The CHSH game value is an average of $J_i$ for all $n$ experimental trials,
\begin{equation}\label{Eq:average}
	\bar{J} = \frac{1}{n}\sum_{j=1}^{n}J_i-3/4.
\end{equation}
Here, we consider the case that the average probability of measurement setting choice is unbiased, $p(xy)=1/4$.
Violating the inequality in Eq. (1), $\bar{J} > 0$, indicates the presence of genuine quantum randomness in the measurement outcomes. The randomness can be quantified by the smooth min-entropy $H_{\min}^{\varepsilon_s}(\textbf{AB}|\textbf{XY}E)$ based on the CHSH game value $\bar{J}$ and the number of experiment trials \cite{arnon2016simple}, which is bounded by,
\begin{equation}\label{Eq:randomness}
	H_{\min}^{\varepsilon_s}(\textbf{AB}|\textbf{XY}E) \ge n\cdot R_{opt}(\varepsilon_s,\varepsilon_{\ea}, \oexp).
\end{equation}
Here $\textbf{A}$ ($\textbf{B}$) and $\textbf{X}$ ($\textbf{Y}$) denote the input and output sequences of Alice (Bob), respectively; $E$ denotes side information of a general quantum adversary; $\varepsilon_s$ is the smoothing parameter;
$\oexp$ is the expected CHSH game value.
$\varepsilon_{\ea}$ is the probability of aborting the protocol; $H_{\min}^{\varepsilon_s}(\textbf{AB}|\textbf{XY}E)$ describes the amount of unpredictable randomness that can be extracted from the outputs $\textbf{AB}$ against the inputs $\textbf{XY}$ and any adversary $E$; as a conservative estimation, we take the lower bound $R_{opt}(\varepsilon_s,\varepsilon_{\ea}, \oexp)$ as the theoretical amount of randomness on average for each trial.
$\varepsilon^c_{\textrm{QRNG}}$ is the \emph{completeness} error, $i.e.$, the probability for a protocol to abort for an honest implementation is at most $\varepsilon^c_{\textrm{QRNG}}$. The lower bound for the smooth min-entropy $H_{\min}^{\varepsilon_s}(\textbf{AB}|\textbf{XY}E)$ is smaller when the average probability of measurement setting choice is biased (See Supplemental Materials).

We experimentally implement the self-testing QRNG with the following procedure,
\begin{enumerate}
	\item Bell test:
	\begin{enumerate}
		\item In experimental trial $i$, Alice and Bob receive random inputs $X_i$ and $Y_i$ and produce outputs $A_i$ and $B_i$, respectively.
		\item We assign a CHSH game value $J_i$ according to the pay-off in Eq.~\eqref{Eq:Jpayoff} and calculate the average pay-off according to Eq.~\eqref{Eq:average}.
		\item We abort the protocol if $\bar{J}< \oexp - \dest $. Here the completeness error is upper bounded by $\varepsilon^c_{\textrm{QRNG}}\le \exp{(-2n\dest^2)}$, where $\delta_{est}\in (0,1)$ is the width of the statistical confidence interval for the Bell violation estimation test.
	\end{enumerate}
	\item Randomness estimation: conditioned on the violation of Bell's inequality in the experiment, either the protocol aborts with probability greater than $1-\varepsilon_{\ea}$ or the experiment produces randomness given by Eq.~\eqref{Eq:randomness}.
	\item Randomness extraction: for a given failure probability of less than $2^{-t_e}$, we apply the Toeplitz-matrix hashing extractor with a matrix of size $n\times H_{\min}^{\varepsilon_s}(\textbf{AB}|\textbf{XY}E) - t_e$ to extract $H_{\min}^{\varepsilon_s}(\textbf{AB}|\textbf{XY}E) - t_e$ random bits that is $\varepsilon_s$ close to the uniform distribution. Here, we set $t_e = 100$.
\end{enumerate}

\begin{figure*}[tbh]
\centering
\resizebox{13cm}{!}{\includegraphics{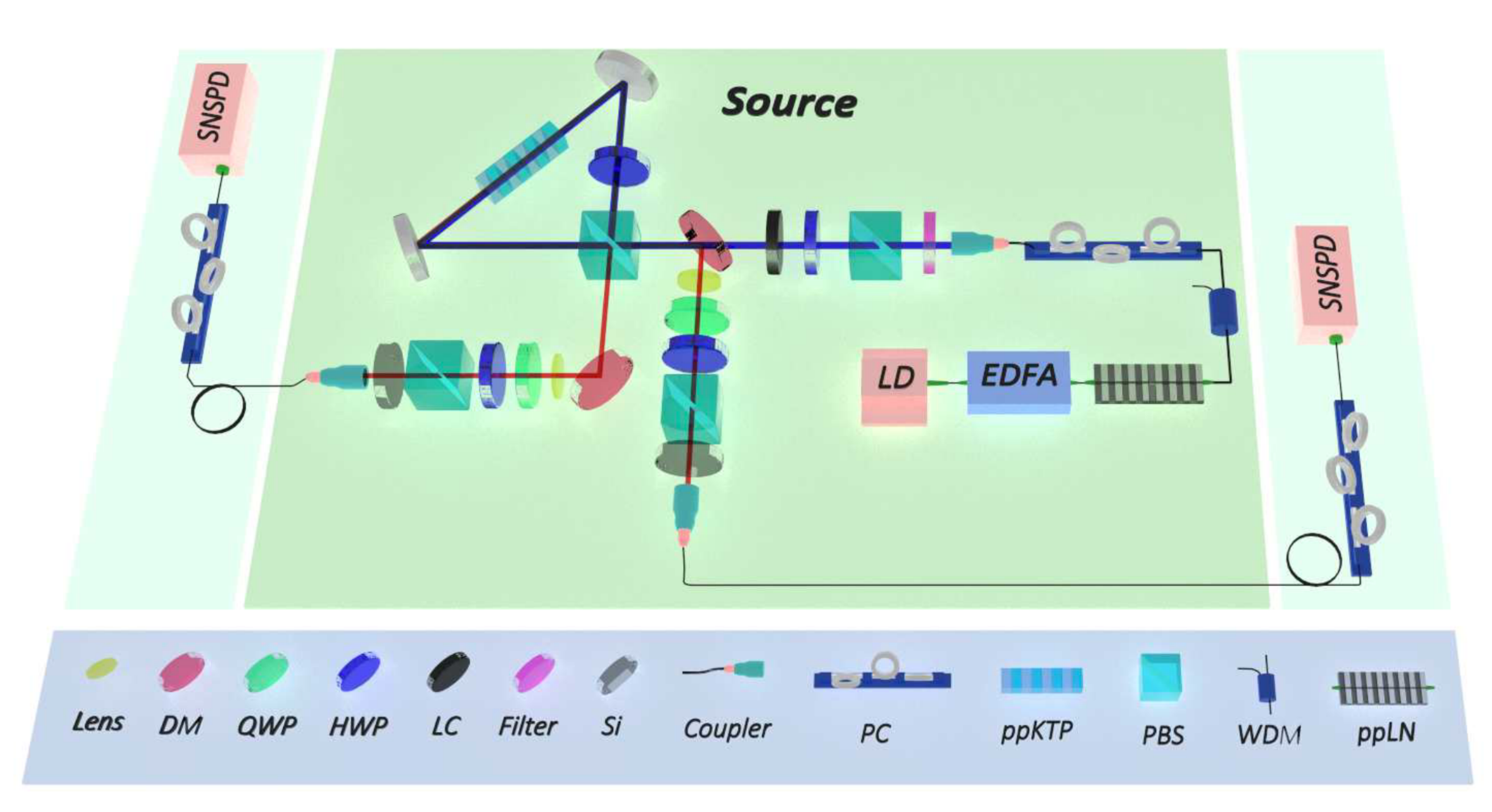}}
\caption{Schematics for self-testing QRNG.
Light pulses of 10 ns, 100 kHz from a 1560 nm seed laser (LD) are amplified by an erbium-doped fiber amplifier (EDFA), and up-converted to pulses at 780 nm via second-harmonic generation (SHG) in an in-line periodically poled lithium niobate (PPLN) waveguide. The residual 1560 nm light is removed by a wavelength-division multiplexer (WDM) and spectral filters. A half-wave plate (HWP) and a liquid crystal (LC) is used to adjust the pump polarization. The 780 nm light pulses are focused into a periodically poled potassium titanyl phosphate (PPKTP) crystal in a Sagnac loop to generate polarization entangled photon pairs. After removing the 780 nm pump pulses by dichroic mirrors (DMs) and a 1 mm thick silicon plate, the entangled photons at 1560 nm are subject to polarization state measurements and then sent to superconducting nanowire single-photon detectors (SNSPD) via 130 m optical fiber.}
\label{Fig:DISetup}
\end{figure*}

{\it Experiment.---}
The experimental layout is shown in Fig.~\ref{Fig:DISetup}. We enclose a periodically poled potassium titanyl phosphate (PPKTP) crystal in a Sagnac loop. With the injection of pump pulses at wavelength of 780 nm and pulse width of 10 ns at a repetition rate of 100 kHz, the loop emits polarization-entangled photon pairs at the degenerate wavelength of 1560 nm via spontaneous parametric downconverison process. The two photons of the pair travel in opposite directions. They are subject to polarization state measurements by Alice and Bob and are detected by superconducting nanowire single-photon detectors (SNSPD). In each experimental trial, the dichotomic photon-detection results of SNSPDs, 1 for "click" and 0 for "no click", are time-tagged for correlation analysis.

In our experiment, the overall single photon detection efficiency is determined to be $78.6\%\pm1.5\%$ for Alice and $80.2\%\pm1.5\%$ for Bob \cite{Pereira2013}. The two-photon interference visibility is measured to be $99.5\%\pm2\%$ ($97.8\%\pm1.5\%$) in the horizontal (diagonal) basis. Using Eberhard's method \cite{Eberhard93}, we generate nonmaximally entangled state, $\cos(20.5^\circ)\ket{HV}+\sin(20.5^\circ)\ket{VH}$ and set angles of half-wave plates  in polarization state measurements to be $A_1=-84.0^\circ$ or $A_2=-118.7^\circ$ for Alice, $B_1=6^\circ$ or $B_2=-28.7^\circ$ for Bob, respectively, to have an optimum violation of Bell's inequality (see Supplemental Materials).

Previous Bell test experiments assigned measurement settings ($xy$) randomly with inputs from locally generated QRNGs \cite{hensen2015loophole, Shalm15, Giustina15, Rosenfeld16}.  RNGs including cosmic RNGs will be exploited to address the no-signaling issue in the input randomness \cite{Gallicchio14,Handsteiner17,cheng2016random}. For the current experiment, the settings are preset manually with parameters given by the Eberhard's optimization procedure as described above. The total photon detection efficiencies are high to close the detection loophole. We repeat the experiment by equal number of trials ($N = 1\times10^{10}$) per measurement setting choice $xy$ and record the number of correlated events $N_{ab|xy}$ (see Tab.~\ref{tab:Eberhard}). According to Eq. (4-6), the $J$-value of the CHSH game is given by
\begin{equation}\label{Eq:expviolation}
	J_N = J_{A_1 B_1}+J_{A_1 B_2}+J_{A_2 B_1}+J_{A_2 B_2}-3/4,
\end{equation}
with the Bell value $J$ for settings $x=A_i$, $y=B_j$, and outputs $ab$ given by
\begin{equation}
	\begin{cases}
	J_{A_1 B_1} = (N_{ab=00|A_1 B_1}+N_{ab=11|A_1 B_1})/N,\\
	J_{A_1 B_2} = (N_{ab=00|A_1 B_2}+N_{ab=11|A_1 B_2})/N,\\
	J_{A_2 B_1} = (N_{ab=00|A_2 B_1}+N_{ab=11|A_2 B_1})/N,\\
	J_{A_2 B_2} = (N_{ab=01|A_2 B_2}+N_{ab=10|A_2 B_2})/N.\\
	\end{cases}
\end{equation}

For a total number of experimental trials $n = 4N = 4 \times 10^{10}$, the obtained $J$-value is $3.52 \times 10^{-4}$, indicating that our CHSH game rejects local hidden variable models. 
In our analysis, we set the expected CHSH game value to the one measured in the experiment, $\oexp = 3.52\times10^{-4}$, $\varepsilon_s = \varepsilon_{\ea} = 1/\sqrt{n}= 5 \times 10^{-6}$ and $\dest = \sqrt{10/n}= 1.58 \times 10^{-5}$.
Correspondingly, after applying an 80 Gb $\times$ 45.6 Mb Toeplitz matrix hashing, the experiment produces $4.56 \times 10^{7}$ genuine random bits in total with uniformity within $\varepsilon_s + \varepsilon_{\ea} =  10^{-5}$. The randomness generation speed is $0.00114$ bits per trial or 114 bits/s, which is an improvement of $\times 10^7$ over the previous experiment \cite{pironio2010}. With such a high yield, the stream of random bits pass the NIST statistic test suite for the first time of its kind (see Supplemental Materials). 
\begin{table}[htb]
\centering
  \caption{Number of correlated events for equal number of trials ($10^{10}$) per measurement base settings: $A_1 B_1$, $A_1 B_2$, $A_2 B_1$ and $A_2 B_2$. $a = 0$ or $1$ indicates that Alice detects a photon or not, the same $b$ for Bob. Mean photon number $\mu=0.15$, violation $J_n=3.52\times10^{-4}$.}
\begin{tabular}{ccccc}
\hline
Basis settings & $ab=00$ & $ab=10$ & $ab=01$ & $ab=11$\\
\hline
$A_1 B_1$ & 9780816728 & 49862593  & 57002217  & 112318462 \\
$A_1 B_2$ & 9574958251 & 41366122  & 263425568 & 120250059 \\
$A_2 B_1$ & 9577555854 & 253683380 & 46874032  & 121886734 \\
$A_2 B_2$ & 9255451323 & 361101111 & 365181363 & 18266203  \\
\hline
\end{tabular}
\label{tab:Eberhard}
\end{table}

\begin{figure}[htb]
  \centering
\resizebox{9cm}{!}{\includegraphics[scale=1]{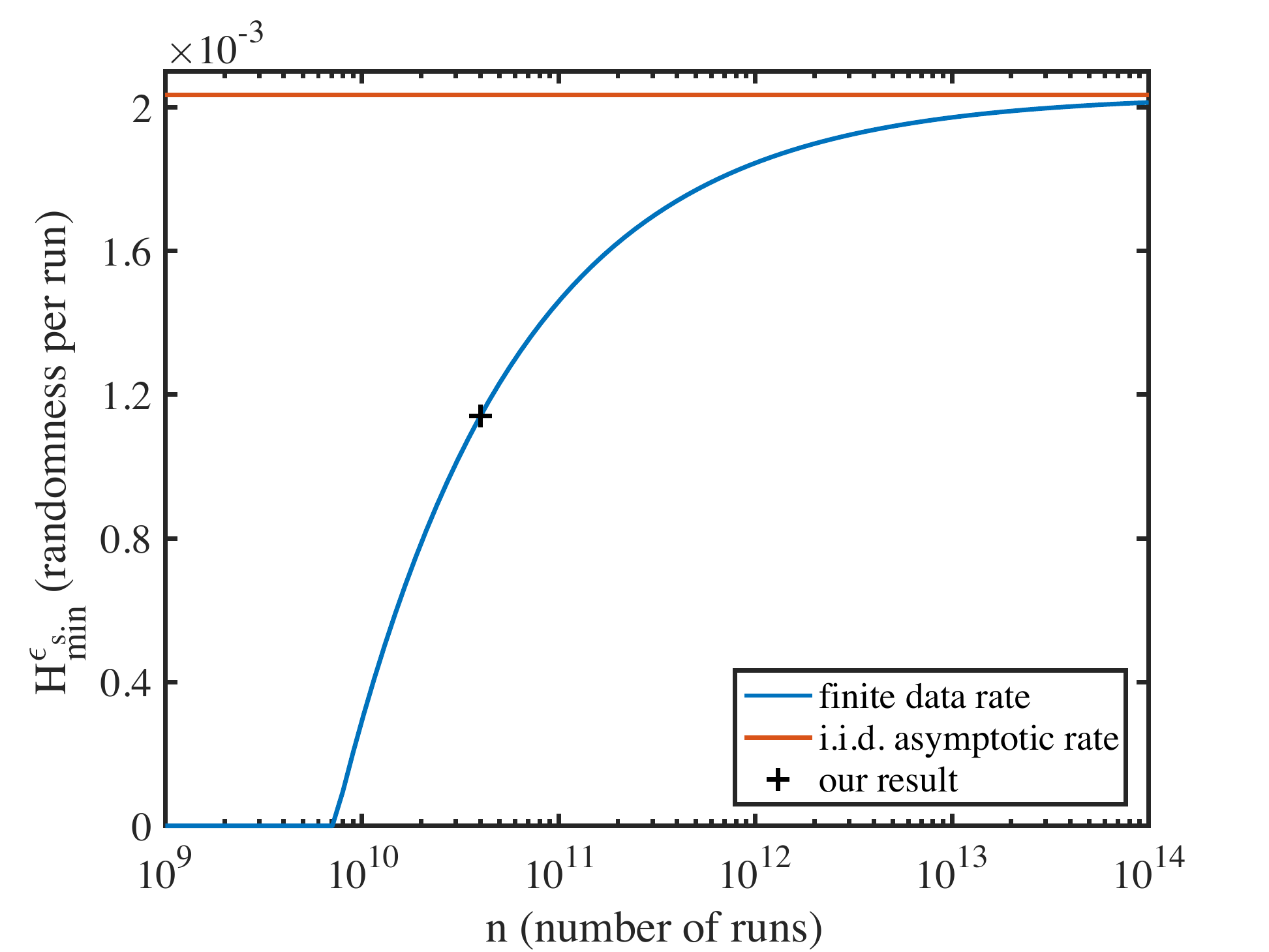}}\\
  \caption{Randomness generation versus number of experimental trials. We set the expected CHSH game value to be $\oexp = 3.52\times10^{-4}$, $\varepsilon_s = \varepsilon_{\ea} = 1/\sqrt{n}$ and $\dest = \sqrt{10/n}$ for finite data rate. }\label{Fig:randomN}
\end{figure}

With the same parameter setting, we plot the amount of randomness that can be produced by our experiment as a function of the number of experimental trials, which asymptotically approaches the optimal asymptotic value for i.i.d. as shown in Fig.~\ref{Fig:randomN}. The amount of randomness obtained in the current experiment is about 60\% of the optimal asymptotic value.

One may expect to extract more randomness, even by orders of magnitude, for larger violations in the CHSH game with improved experimental parameters such as higher photon-detection efficiency, higher two-photon interference visibility, and optimized mean photon numbers (see Supplemental Materials), as shown in Fig.~\ref{Fig:randomViolation.eps}.

\begin{figure}[htb]
  \centering
  \resizebox{9cm}{!}{\includegraphics[scale=1]{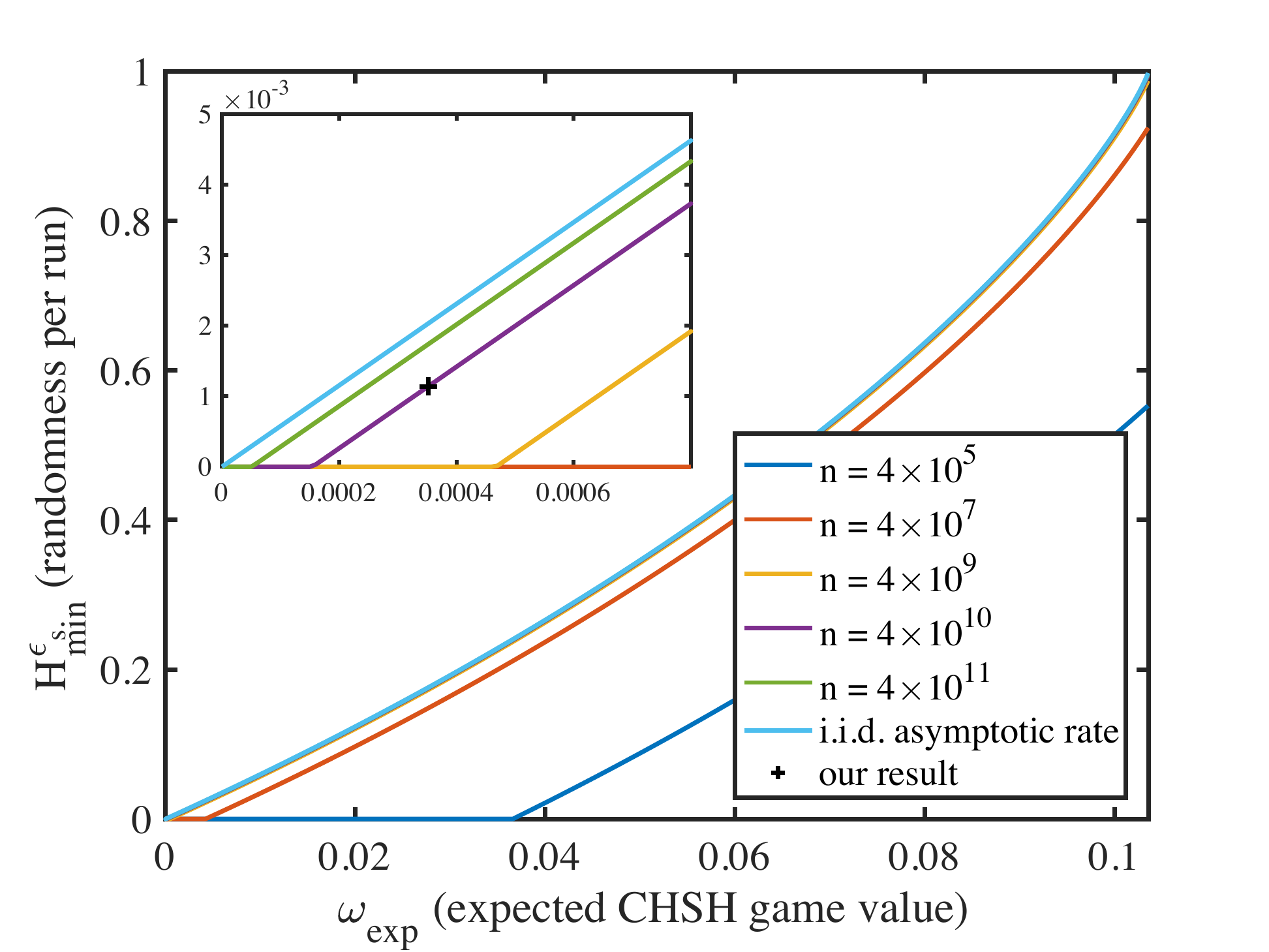}}\\
  \caption{Randomness generation versus expected CHSH game value. We set $\varepsilon_s = \varepsilon_{\ea} = 1/\sqrt{n}$ and $\dest = \sqrt{10/n}$ for finite data rate curves. }\label{Fig:randomViolation.eps}
\end{figure}

{\it Conclusion and outlook.---} We implement a self-testing QRNG without the detection loophole, which does not need the assumption of identical and independent distribution and is against quantum adversary. 
We report randomness extraction at a rate of 114 bits/s with uniformity within $10^{-5}$, marking a critical step in generating self-testing QRNG for secure information processing applications and tests of fundamental physics. One of the prospective works is to employ cosmic RNGs in a loophole free CHSH game such that the generated random bits are genuinely quantum-certified. Also, one can further speedup the randomness production by upgrading the utilized technology such as scaling up the operation repetition rate.
One may note that the analysis method in Ref.~\cite{arnon2016simple} is efficient in producing quantum-certified randomness based on a CHSH game. However, it remains an open question what is the actual maximum extractable randomness in a CHSH game, furthermore, what is the maximum extractable randomness for a general nonlocal game. 

{\it Note added.---}  Recently, we become aware of a related work \cite{bierhorst2017experimentally}.

\emph{Acknowledgement.}---The authors would like to thank Y. Shi for enlightening discussions. This work has been supported by the National Fundamental Research Program (under Grant No. 2013CB336800), the National Natural Science Foundation of China, the Chinese Academy of Science, and the 1000 Youth Fellowship program in China.

\section*{Supplemental Material}
\section{Theory of self-testing quantum random number generation}
Here, we consider self-testing random number generation based on Bell test. Given a randomness source $R$ as input, we can expand the input randomness by exploiting the protocol described below. At the end of the protocol, a classical string $Z$ is returned. Conditioned on the random source $R$ and a potential adversary's system $E$, the classical string $Z$ is close to uniform. Note that $E$ may be initially correlated to the implementation device. While running the protocol, the device is in the control of user and cannot signal to Eve.
Before introducing the protocol, we define the soundness $\varepsilon^{s}_{QRNG}$ and completeness $\varepsilon^{c}_{QRNG}$ errors of a self-testing randomness generation protocol.

\begin{definition} A randomness expansion protocol is $(\varepsilon^{c}_{QRNG}, \varepsilon^{s}_{QRNG})$-secure, if it satisfies:
\begin{enumerate}
\item Soundness: For an untrusted device, it either aborts with probability less than $P_{abort}$ or returns a string $Z$ satisfying
    \begin{equation}
    (1-P_{abort})\|\rho_{ZRE}-\rho_{U_m}\otimes \rho_{U_r}\otimes \rho_{E}  \|\le \varepsilon^{s}_{QRNG}.
    \end{equation}
    Here, $\rho_{ZRE}$ denotes the state of the classical string $Z$, the input randomness source $R$ and the system of Eve $E$; $\rho_{U_r}$ denotes a string of uniform random bits. The above inequality indicates that the output string is almost ($\varepsilon^{s}_{QRNG}$) a uniform bit string and it has no correlation with the input randomness and the adversary Eve.
\item Completeness: The honest device does not abort with probability greater than $1-\varepsilon^{c}_{QRNG}$.
\end{enumerate}
\end{definition}

\subsection{Protocol description}
The inputs for the Bell test experiment are required to be random, which consumes random numbers.
This inefficiency motivates the spot-checking protocol \cite{Coudron2013,Miller15, arnon2016simple}, where Bell test is only run with probability $q$ (usually small) and the input is fixed with probability ($1-q$).
A CHSH spot-checking protocol is shown as follows.
\begin{enumerate}
	\item Bell test:
\begin{enumerate}
\item
The game involves two players, Alice and Bob. Each player inputs a bit and is required to output a bit in each trial, for a total number of $n$ trials.  Denote their input and output bits in trial $i$ to be $X_i$ and $Y_i$ and $A_i$ and $B_i$, respectively.
\item
Choose a bit  string $\mathbf{t}=(t_1,\cdots,t_n)$, for any $i\in \{1,2,\cdots,n\}$, $ t_i\in\{0,1\}$ according to the distribution $(1-q,q)$.
\item
If $t_i=1$, this trial is a test trial, which is used to test the existence of adversary. Then Alice and Bob randomly input $X_i=\{0,1\}$, $Y_i=\{0,1\}$ and obtain $A_i=\{0,1\}$, $B_i=\{0,1\}$. In each trial, a score is recorded according to a pay-off function $J_i(A_iB_iX_iY_i)\in \{0,1\}$, which is given as follows.
\begin{equation}
J_i(A_iB_iX_iY_i) = \begin{cases}1 &A_i+B_i = X_i*Y_i \\ 0 & A_i+B_i \neq X_i*Y_i\end{cases}
\end{equation}
\item
If $t_i=0$, this trial is a generation trial with fixed input string, $X_i=0$ $Y_i=0$ and $J_i=0$.
\item
(c)-(d) steps are repeated in $n$ trials.
\item
If $\sum_i J_i/n - 0.75< \oexp q - \dest$, we abort the protocol. Otherwise, the protocol can be used for quantum random number generation. And the completeness error is upper bounded by $\varepsilon^c_{QRNG}\le \exp{(-2n\dest^2)}$.
\end{enumerate}
\item Randomness estimation: conditioned on the Bell test is not aborted, either the protocol aborts with probability less than $\varepsilon^c_{QRNG}$ or the amount of randomness is given by Eq.~\eqref{Eq:rawkey}.
\item Randomness extraction: with failure probability less than $2^{-t_e}$, we can extract $n\cdot R_{opt}(\varepsilon_s,\varepsilon_{EA}) - t_e$ random bits that is $\varepsilon^s_{QRNG}$ close to a uniform distribution by using the Toeplitz-matrix hashing matrix.
\end{enumerate}

\subsection{Estimation of randomness production}

In this work, we mainly focus on the security proof in Ref.~\cite{arnon2016simple} . 
The main tool of the security proof is based on the entropy accumulation
theorem (EAT) \cite{dupuis2016entropy}, which can offer an avenue to reduce the complex multi-trial protocol to the i.i.d. case.
After proving that the device independent protocol satisfies the requirements of an EAT channel, the entropy generated in the protocol can be analyzed based on the EAT theorem. Next, the randomness production based on the violation of CHSH inequality \cite{pironio2009device} under the stronger i.i.d. assumption can be connect to the non-i.i.d. case and gives a good lower bound for the generated randomness. 

According to Theorem 10 of \cite{arnon2016simple}, the optimal randomness yield is given by 
\begin{equation}\label{Eq:rawkey}
H_{min}^{\varepsilon_s}(\textbf{AB}|\textbf{XY}E)
 \ge  n \cdot R_{opt}(\varepsilon_s, \varepsilon_{EA}).
\end{equation}
where the smoothed min-entropy $H_{min}^{\varepsilon_s}(\textbf{AB}|\textbf{XY}E)$ evaluates the amount of extractable randomness, $\varepsilon_s$ is the smoothing parameter, $\varepsilon_{EA}$ is the error probability of the entropy accumulation protocol.
Here, $R_{opt}(\varepsilon_s, \varepsilon_{EA})$ is defined as (given fixed $q, n$)
\begin{align}
g(p) = \begin{cases}1-h\left(\frac{1}{2}+\frac{1}{2}\sqrt{16\frac{p}{q}(\frac{p}{q}-1)+3}\right)& \frac{p}{q} \in [0,\frac{2+\sqrt{2}}{4}]\\ 1 & \frac{p}{q} \in [\frac{2+\sqrt{2}}{4},1]\end{cases}
\end{align}
\begin{align}
f_{\min}(p,p_t) = \begin{cases}g(p)&p\leq p_t\\ \frac{d}{dp}g(p)|_{p_t}\cdot p + (g(p_t) - \frac{d}{dp}g(p)|_{p_t}\cdot p_t) & p > p_t\end{cases}
\end{align}
\begin{align}
&R(p, p_t, \varepsilon_s, \varepsilon_e)\\
 &= f_{\min}(p, p_t) - \frac{1}{\sqrt{n}}2(\log 13 + \frac{d}{dp}g(p)|_{p_t})
\sqrt{1-2\log (\varepsilon_s \cdot \varepsilon_e)}.
\end{align}
\begin{align}
R_{opt}(\varepsilon_s, \varepsilon_e) = \max_{\frac{3}{4}<\frac{p_t}{q}<\frac{2+\sqrt{2}}{4}} R(\omega_{exp}\cdot q - \delta_{est}, p_t, \varepsilon_s, \varepsilon_e).
\end{align}
where $\omega_{exp}$ is the expected winning probability for an honest but noisy device, $\delta_{est}\in (0,1)$ is the width of the statistical confidence interval
for the Bell violation estimation test. 
Thus $\omega_{exp}$ can be chosen according to the experimental winning scores $\sum_i J_i$ and $\omega_{exp}=\sum_i J_i/n - 0.75$. A greater $\delta_{est}$ will lead to a higher randomness production. However, $\delta_{est}$ is bounded by the number of trials $N$ and another error $\varepsilon^{c}_{QRNG}$, which represents the completeness error of the entropy accumulation protocol,
\begin{equation}
\varepsilon^{c}_{QRNG}\le \exp(-2n \delta_{est}^2)
\end{equation}
Considering the failure probability $2^{t_e}$ in the randomness extraction process, the soundness error for this protocol is 
\begin{equation}
\varepsilon^{s}_{QRNG}=\varepsilon_s+ \varepsilon_{EA} +2^{t_e}.	
\end{equation}

\subsection{Optimal parameter $q$}
Define the net randomness output to be the difference between the output and input randomness. In general, we can optimize the parameter $q$ to find out the maximal net randomness output per round,
\begin{equation}
R_{net} =  R_{opt}(\varepsilon_s, \varepsilon_{EA}) -h(q)-q
\end{equation}
For example consider $\delta_{est}=10^{-2}, \varepsilon_s = 10^{-6}, \varepsilon_{EA} = 10^{-6}, w_{exp} = \frac{2+\sqrt{2}}{4}-0.75$, the optimal $q$ which maximizes the $R_{net}$ can be found by numerical search for a given $n$. Here we give a table for different $n$ and the corresponding optimal $q$ and $R_{net}$.
\begin{table}
\caption{The relation between the net randomness rate, bias $q$, and number of experimental trials $n$.}
\begin{tabular}{|c|c|c|}
\hline
$n$ & $q$ & $R_{net}$\\
\hline
$10^{7}$ &- & 0\\
\hline
$10^{8}$ & 0.0777& 0.0256\\
\hline
$10^{9}$ & 0.0587& 0.2175\\
\hline
$10^{11}$ & 0.0486& 0.3279\\
\hline
$10^{14}$ & 0.0473 & 0.3417\\
\hline
\end{tabular}
\end{table}

In our experiment, we set $q = 1$. As we do not use randomness to choose the measurement setting, the obtained randomness $nR_{opt}(\varepsilon_s, \varepsilon_{EA})$ is the net randomness output.

\subsection{Numerical result}
\begin{figure}[htb]
  \centering
  \resizebox{9cm}{!}{\includegraphics[scale=1]{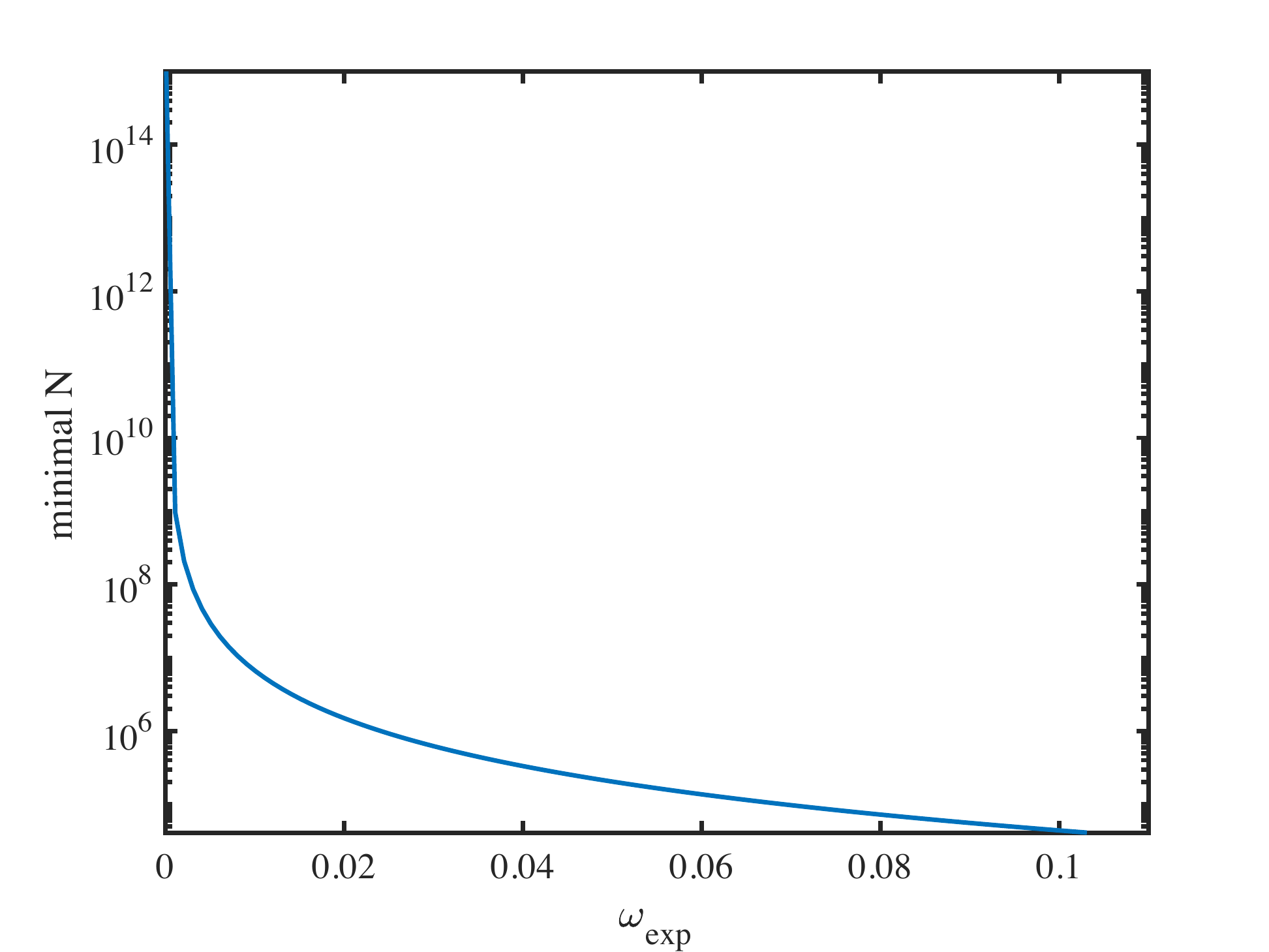}}\\
  \caption{Minimal number of experimental trials $n$ that has a nonzero randomness generation rate for different expected violations. Here, we set $\varepsilon_s = \varepsilon_{\ea} = 1/\sqrt{n}$ and $\dest = \sqrt{10/n}$ for different number of experimental trials $n$. }\label{Fig:minmalN.pdf}
\end{figure}

As shown in Fig.~\ref{Fig:randomViolation.eps}, we consider the relation between the randomness output $R_{opt}(\varepsilon_s, \varepsilon_{EA})$ and the expected CHSH value $\omega_{exp}$. Given the expected CHSH value $\omega_{exp}$, we also show in Fig.~\ref{Fig:minmalN.pdf} the minimal required number of trials to guarantee a nonzero randomness output. 

\begin{figure}[htb]
  \centering
  \resizebox{9cm}{!}{\includegraphics[scale=1]{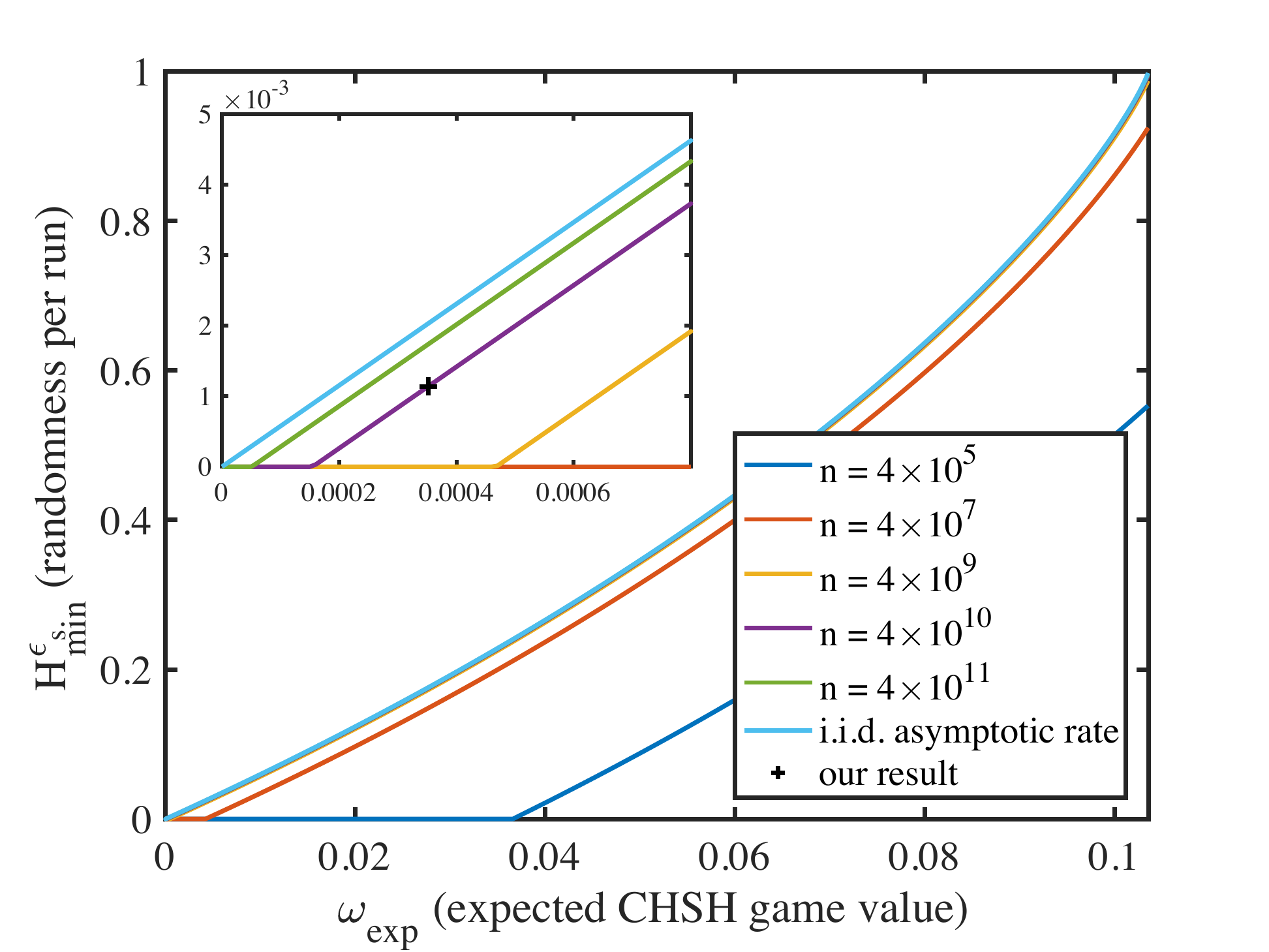}}\\
  \caption{Randomness generation with different expected violation. For the finite data rate curve, we set $\varepsilon_s = \varepsilon_{\ea} = 1/\sqrt{n}$ and $\dest = \sqrt{10/n}$ for different number of experimental trials $n$. }\label{Fig:randomViolation.eps}
\end{figure}

\begin{figure}[htb]
  \centering
  \resizebox{9cm}{!}{\includegraphics[scale=1]{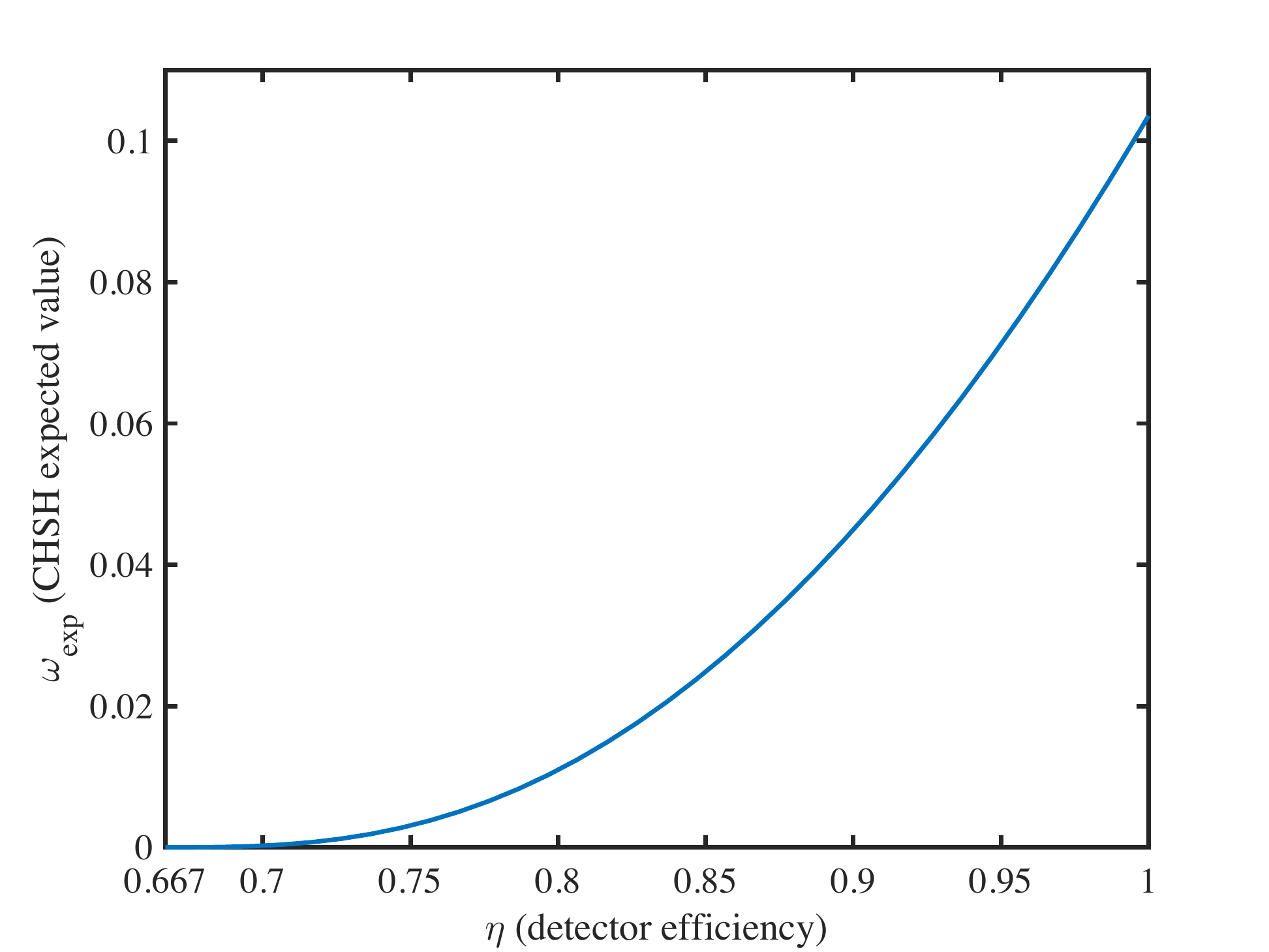}}\\
  \caption{Maximal expected CHSH value $\omega_{exp}$ with different detection efficiency, considering the perfect state preparation and imperfect detection with efficiency $\eta$.}\label{Fig:efficiency_randomness.pdf}
\end{figure}

\begin{figure}[htb]
  \centering
  \resizebox{9cm}{!}{\includegraphics[scale=1]{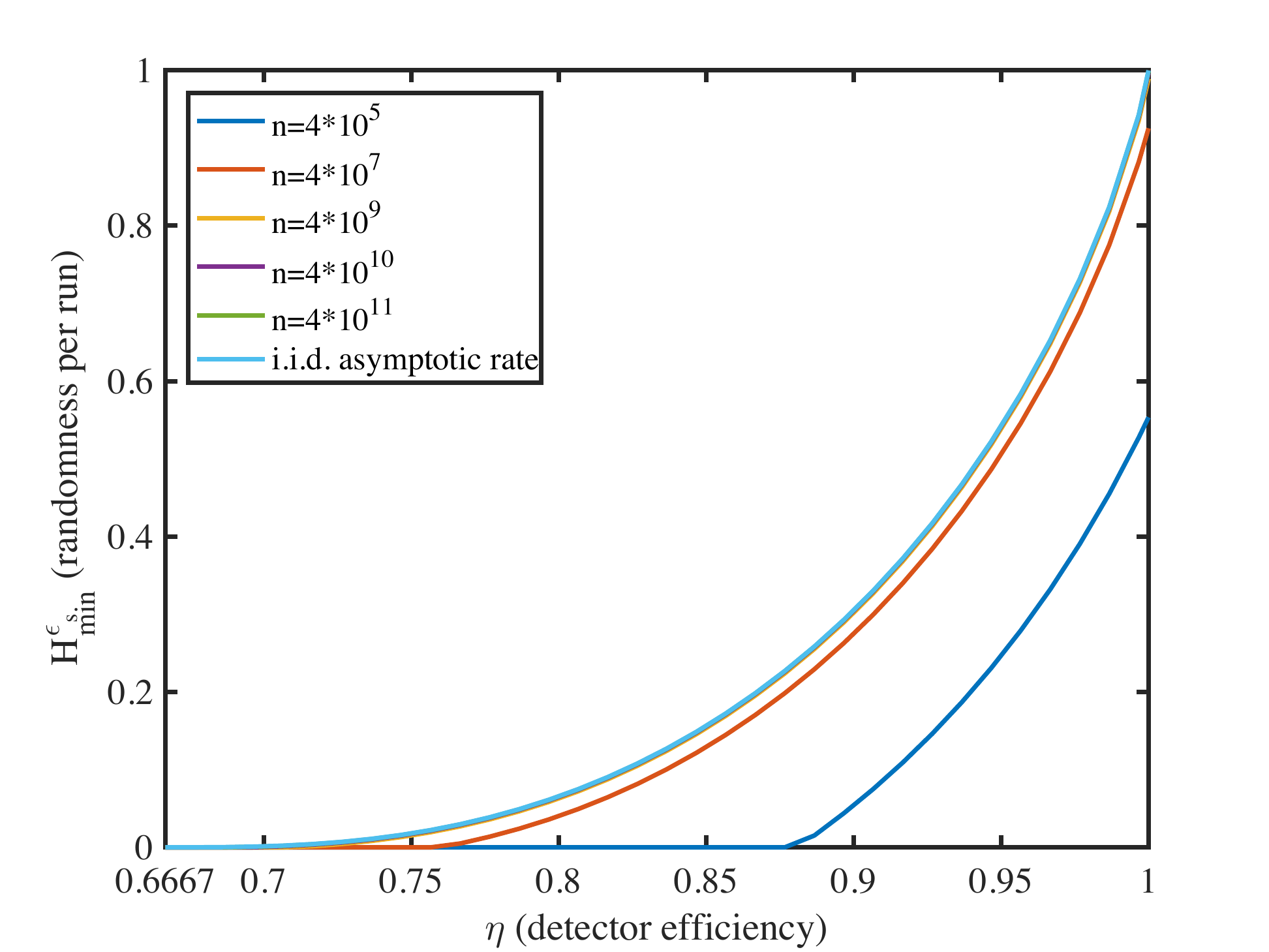}}\\
  \caption{Maximal randomness generation with different detection efficiency, considering the perfect state preparation and imperfect detection with efficiency $\eta$. For the finite data rate curve, we set $\epsilon_s = \epsilon_{\ea} = 1/\sqrt{n}$ and $\dest = \sqrt{10/n}$ for different number of runs $n$. }\label{Fig:efficiency_violation.pdf}
\end{figure}

Considering perfect state preparation and imperfect detection with efficiency $\eta$, the relation between the maximal expected CHSH value $\omega_{exp}$ and the efficiency is shown in Fig.~\ref{Fig:efficiency_randomness.pdf}. We also show the relation between the detection efficiency and the randomness output in Fig.~\ref{Fig:efficiency_violation.pdf}.

\section{Entanglement characterization}
To characterize the entangled photon pairs, we set the mean photon number to $\mu=0.004$ to decrease the multi-photon effect. We set the pump to be diagonal polarized $\ket{+}$, to generate the maximum entangled state $\ket{\Psi^+}=(\ket{HV}+\ket{VH)/\sqrt2}$. The visibility is measured under horizontal/vertical basis and diagonal/anti-diagonal basis, as $99.5\pm 2.0\%$ and $97.8\pm 1.5\%$. We attribute the imperfection to multi-photon components, imperfect optical elements, and imperfect spatial/spectral mode matching. 

In testing the Bell inequality, we need to generate a non-maximally entangled state. Considering the system heralding efficiency and the visibility, we optimize the state and measurement basis based on $77\%$ system efficiency, following Eberhard's original article \cite{Eberhard93}. The sub-optimal state is $\cos(20.5^\circ)\ket{HV}+\sin(20.5^\circ)\ket{VH}$ (with $r=0.37$ for $(\ket{HV}+r\ket{VH})/\sqrt{1+r^2}$). The measurement bases are  $a_1=-84.0^\circ$, $a_2=-118.7^\circ$ for Alice, $b_1=6^\circ$, $b_2=-28.7^\circ$ for Bob.

\begin{figure}[htb]
\centering
    \subfigure[]{
      \includegraphics[width=8cm]{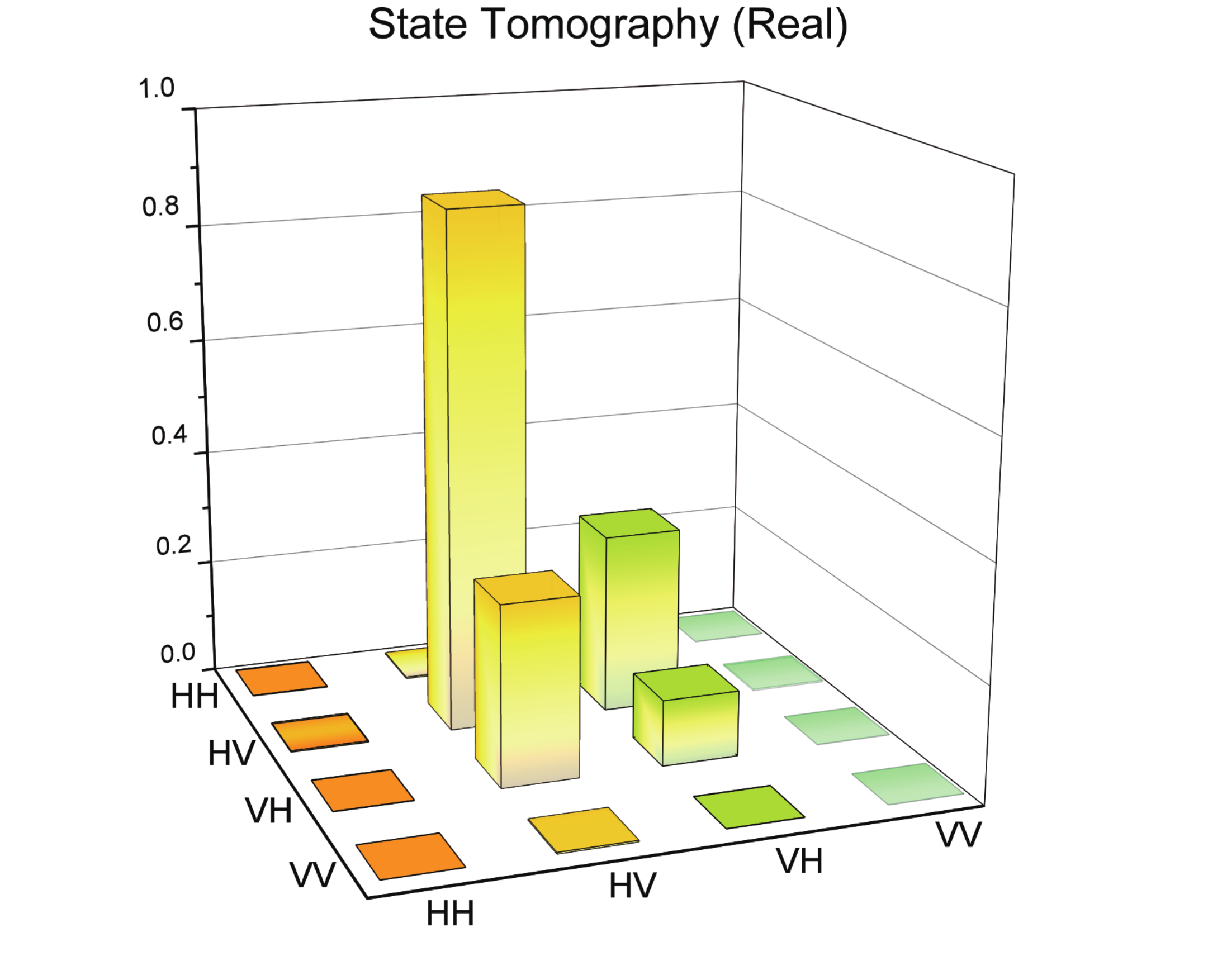}
    }
    \subfigure[]{
      \includegraphics[width=8cm]{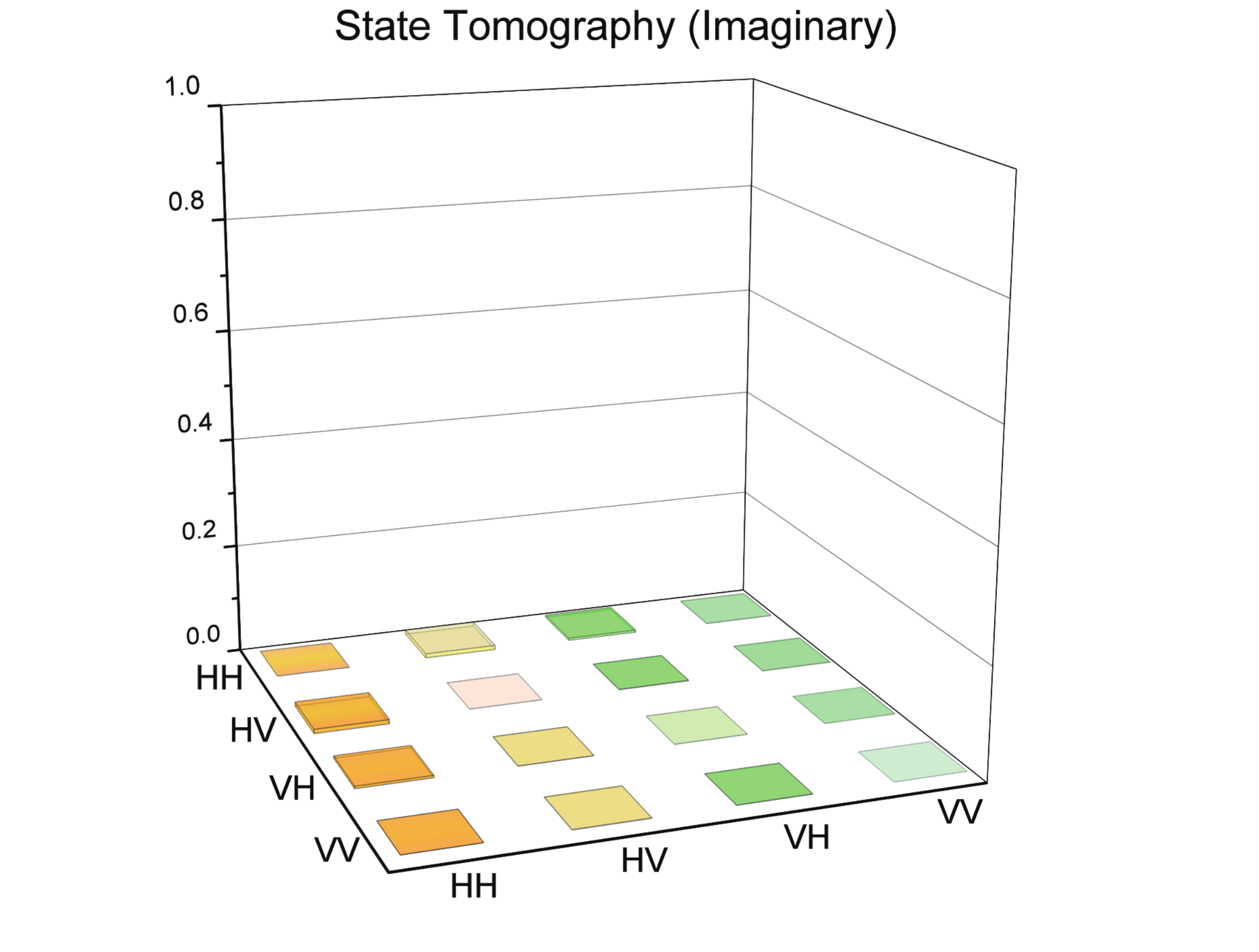}
    }
\caption{(color online)
Tomography of the produced state. The real and imaginary part are shown in (a) and (b), respectively.}
\label{Fig.Tomo}
\end{figure}

We perform state tomography on the non-maximally entangled state $\cos(20.5^\circ)\ket{HV}+\sin(20.5^\circ)\ket{VH}$. The result is shown in Fig.~\ref{Fig.Tomo}, the state fidelity is $99.3\%$ compared to the ideal state.

\section{System efficiency}
The system efficiency is vital to close the detection loophole in Bell test. Normally, the system efficiency includes the following parts: the entanglement source, the transmission link, the detection. Here we consider each part individually.

The (system) heralding efficiency is defined as $\eta_s=C/N_i$ and $\eta_i=C/N_s$, where the coincidence events (C) and the signal (idler) events $N_s$($N_i$) can be measured during the experiment. If considering the optical efficiency $\eta_{opt}$ and the detector efficiency $\eta_s^{det}$($\eta_i^{det}$), the entangled pair collection efficiency is $\eta_s^{opt}=C/(N_i \cdot \eta_s^{det} \cdot \eta_{opt})$ and $\eta_i^{opt}=C/(N_s \cdot \eta_i^{det} \cdot \eta_{opt})$.

In a Sagnac based entanglement generation as we used in the experiment, a periodically poled potassium titanyl phosphate (PPKTP) crystal is used as the nonlinear material to generate photon pairs. The generated photon pairs will be collected into single mode fibers for further modulation and detection. Theory predictions \cite{Bennink2010} and experimental tests \cite{Pereira2013, Dixon2014} show it is possible to achieve a near unity heralding efficiency by selecting appropriate focal parameters. For our source, we adjust the pump light beam waist to be around 180 $\mu m$, and the collection beam waist to be around 85 $\mu m$ (see the setup section for detail). Theory predicts the collection efficiency for the photons is 98\% if the PPKTP is 0.9 cm thick.

In detection, We uses high efficiency superconducting nanowire single-photon detectors (SNSPD) \cite{Zhang_NbN_2016} to detect the received photons. The SNSPD is made of polycrystalline NbN, and can achieve  its best performance (saturate) at around 1.8 K temperature. In our experiment, we cool the detector chips to less than 1K. In our test, the detectors are biased 0.2 $\mu$A below their critical current for stable operation. The single photon detection efficiencies are $90.2\%$ for Alice and $92.2\%$ for Bob, with dark counts less than 100 $s^{-1}$.

We measure the (system) heralding efficiency to be $\eta_s=78.6\%\pm1.5\%$ for Alice and $\eta_i=80.2\%\pm1.5\%$ for Bob. The optical efficiency in the entanglement preparation is around $95.5\%$, the efficiency of the 130 meter fiber between the source and the detection is around $99\%$. We then estimate the entangled pair collection efficiency as $\eta_s^{opt}=92.1\%$ for Alice and $\eta_i^{opt}=92.0\%$ for Bob. 
\section{Detailed Experimental Parameters}
We focus the pump and collection mode to the center of the PPKTP crystal. In experiment, the pump light is spatial filtered by a 780HP single mode fiber. The light is the focused using a aspherical lens with f=8 mm to the PPKTP that is around 70 cm away. The pump waist is estimated to be 180 $\mu$m, and the beam quality $M^2$ is around 1.05.

For collection, we used SMF28e single mode fiber. An Aspherical lens with f=11 mm and a spherical lens with f=175 mm focus the beam to 85 $\mu$m at the center of PPKTP. The spherical lens is placed in the middle, about 19 cm from the aspherical, and about 45 cm from the PPKTP. 

We measure the efficiencies of the optical elements individually, as shown in Tab.~\ref{tab:OptEff}.

\begin{table}[htb]
\centering
  \caption{The efficiencies of optical elements}
\begin{tabular}{c|c}
\hline
Optical element & Efficiency\\
\hline
Asypherical lens & $99.27\%\pm0.03\%$ \\
Spherical lens & $99.6\%\pm1.0\%$ \\
Half wave plate (780nm/1560nm) & $99.93\%\pm0.02\%$ \\
Half wave plate (1560nm) & $99.92\%\pm0.04\%$ \\
Quater wave plate (1560nm) & $99.99\%\pm0.08\%$ \\
Polarization beam splitter (780nm/1560nm) & $99.6\%\pm0.1\%$ \\
Polarization beam splitter (1560nm) & $99.6\%\pm0.2\%$ \\
Silicon plate &	$99.7\%\pm0.2\%$ \\
Dichromatic mirror & $99.46\%\pm0.03\%$ \\
PPKTP & $99.6\%\pm0.2\%$ \\
\hline
\end{tabular}
\label{tab:OptEff}
\end{table}

The dimension of PPKTP we used is 1 mm$\times$2 mm$\times$10 mm, with a poling period of 46.5 $\mu$m.
In our experiment, the pump power is about $3.5$ mw, to create 15000 pairs entangled photons per second.

\section{The bell inequality violation with different mean photon number}

In the experiment, we set the mean photon number to be around 0.15 per pulse. As shown in Tab.~\ref{tab:violation0.15}, the violation is $J_n=3.52\times10^{-4}$ for $4\times10^{10}$ trials.
While previous experiments used  small mean photon number to reduce the multi-photon effect, we find an optimized mean photon number by modeling the Bell test with multi-photons. When the mean photon number is small, the quantum states in most of the experimental trials are vacuum. Increasing mean photon number will increase the proportion of non vacuum states and increase the Bell violation. The violation will decrease when the multi-photon effect becomes significant. We find the optimal violation by numerical simulation, the results are given in Tab.~III-VIII.

\begin{table}[htb]
\centering
  \caption{The CHSH violation with different mean photon numbers.}
\begin{tabular}{c|c}
\hline
Mean photon number & CHSH violation\\
\hline
0.005 & $1.58\times10^{-5}$ \\
0.01  & $6.29\times10^{-5}$ \\
0.02  &	$9.72\times10^{-5}$ \\
0.05  &	$1.99\times10^{-4}$ \\
0.1   &	$2.50\times10^{-4}$ \\
0.15  &	$3.52\times10^{-4}$ \\
0.2	  & $2.48\times10^{-4}$ \\
\hline
\end{tabular}
\label{tab:vioSum}
\end{table}

\begin{table}[htb]
\centering
  \caption{Number of correlated events for equal number of trials ($10^{10}$) per measurement base settings: $A_1 B_1$, $A_1 B_2$, $A_2 B_1$ and $A_2 B_2$. $a = 0$ or $1$ indicates that Alice detects a photon or not, the same $b$ for Bob. Mean photon number $\mu=0.15$, violation $J_n=3.52\times10^{-4}$.}
\begin{tabular}{ccccc}
\hline
Basis settings & $ab=00$ & $ab=10$ & $ab=01$ & $ab=11$\\
\hline
$A_1 B_1$ & 9780816728 & 49862593  & 57002217  & 112318462 \\
$A_1 B_2$ & 9574958251 & 41366122  & 263425568 & 120250059 \\
$A_2 B_1$ & 9577555854 & 253683380 & 46874032  & 121886734 \\
$A_2 B_2$ & 9255451323 & 361101111 & 365181363 & 18266203  \\
\hline
\end{tabular}
\label{tab:violation0.15}
\end{table}

\begin{table}[htb]
\centering
  \caption{Number of correlated events for equal number of trials ($5\times10^{7}$) per measurement base settings. Mean photon number $\mu=0.2$, violation $J_n=2.48\times10^{-4}$.}
\begin{tabular}{ccccc}
\hline
Basis settings & $ab=00$ & $ab=10$ & $ab=01$ & $ab=11$\\
\hline
$A_1 B_1$ & 48453853 & 355764 & 397898 & 792485\\
$A_1 B_2$ & 46955670 & 294787 & 1884426 & 865117 \\
$A_2 B_1$ & 46947701 & 1847449 & 331927 & 872923\\
$A_2 B_2$ & 44662839 & 2563681 & 2598180 & 175300\\
\hline
\end{tabular}
\label{tab:violation0.2}
\end{table}

\begin{table}[htb]
\centering
  \caption{Number of correlated events for equal number of trials ($5\times10^{7}$) per measurement base settings. Mean photon number $\mu=0.1$, violation $J_n=2.50\times10^{-4}$.}
\begin{tabular}{ccccc}
\hline
Basis settings & $ab=00$ & $ab=10$ & $ab=01$ & $ab=11$\\
\hline
$A_1 B_1$ & 49312617 & 155809 & 185673 & 345901\\
$A_1 B_2$ & 48633646 & 127616 & 867571 & 371167 \\
$A_2 B_1$ & 48652215 & 825815 & 149637 & 372333\\
$A_2 B_2$ & 47597903 & 1165406 & 1196733 & 39958\\
\hline
\end{tabular}
\label{tab:violation0.1}
\end{table}

\begin{table}[htb]
\centering
  \caption{Number of correlated events for equal number of trials ($5\times10^{7}$) per measurement base settings. Mean photon number $\mu=0.05$, violation $J_n=1.99\times10^{-4}$.}
\begin{tabular}{ccccc}
\hline
Basis settings & $ab=00$ & $ab=10$ & $ab=01$ & $ab=11$\\
\hline
$A_1 B_1$ & 49657069 & 78257  & 92401  & 172273 \\
$A_1 B_2$ & 49318372 & 65894  & 429597 & 186137 \\
$A_2 B_1$ & 49321033 & 413099 & 78382  & 187486 \\
$A_2 B_2$ & 48789537 & 591262 & 606143 & 13058 \\
\hline
\end{tabular}
\label{tab:violation0.05}
\end{table}

\begin{table}[htb]
\centering
  \caption{Number of correlated events for equal number of trials ($5\times10^{7}$) per measurement base settings. Mean photon number $\mu=0.02$, violation $J_n=9.72\times10^{-5}$.}
\begin{tabular}{ccccc}
\hline
Basis settings & $ab=00$ & $ab=10$ & $ab=01$ & $ab=11$\\
\hline
$A_1 B_1$ & 49848404 & 36312  & 39931  & 75353 \\
$A_1 B_2$ & 49692655 & 31079  & 192707 & 83559 \\
$A_2 B_1$ & 49686075 & 192384 & 35158  & 86383 \\
$A_2 B_2$ & 49448963 & 272187 & 274815 & 4035 \\
\hline
\end{tabular}
\label{tab:violation0.02}
\end{table}

\begin{table}[htb]
\centering
  \caption{Number of correlated events for equal number of trials ($1\times10^{8}$) per measurement base settings. Mean photon number $\mu=0.01$, violation $J_n=6.29\times10^{-5}$.}
\begin{tabular}{ccccc}
\hline
Basis settings & $ab=00$ & $ab=10$ & $ab=01$ & $ab=11$\\
\hline
$A_1 B_1$ & 99837479 & 38578  & 42882  & 81061 \\
$A_1 B_2$ & 99671930 & 33487  & 205544 & 89039 \\
$A_2 B_1$ & 99651185 & 213872 & 40318  & 94625 \\
$A_2 B_2$ & 99396430 & 298159 & 301665 & 3746  \\
\hline
\end{tabular}
\label{tab:violation0.01}
\end{table}

\begin{table}[htb]
\centering
  \caption{Number of correlated events for equal number of trials ($3.58\times10^{8}$) per measurement base settings. Mean photon number $\mu=0.005$, violation $J_n=3.30\times10^{-5}$.}
\begin{tabular}{ccccc}
\hline
Basis settings & $ab=00$ & $ab=10$ & $ab=01$ & $ab=11$\\
\hline
$A_1 B_1$ & 357726090 & 65895  & 74068  & 133947 \\
$A_1 B_2$ & 357459604 & 57387  & 338791 & 144218 \\
$A_2 B_1$ & 357495160 & 309855 & 58769  & 136216 \\
$A_2 B_2$ & 357043095 & 476212 & 475809 & 4884   \\
\hline
\end{tabular}
\label{tab:violation0.005}
\end{table}

\section{Simulation}
In our simulation, we consider the i.i.d. case. We will only focus on the probability distribution. Furthermore, we consider that the experiment observes the no-signaling theorem. In this case, we can equivalently consider the CH inequality.
The CH inequality is a bipartite Bell inequality with measurement settings $x\in\{0, 1\}$ and $y\in\{0, 1\}$, and outputs $a,b\in\{0,1,u\}$, corresponding to 0, 1, and undetected events, respectively. The CH inequality is a linear combination of the probability distribution $p(a,b|x,y)$ and given by
\begin{equation}\label{eq:CHdef}
\begin{aligned}
  J_{CH} &= -p_{00}(0, 0)  -p_{00}(0, 1) -p_{00}(1, 0)\\
  & +p_{00}(1, 1)+ p_0^A(0)  +p_0^B(0)  \geq 0.
   \end{aligned}
\end{equation}
Here, we take a more convenient notation of $p(a,b|x,y)$ as $p_{ab}(x,y)$, and $p_0^A(0)$ ($p_0^B(0)$) refers to the probability of obtaining $0$ when Alice's (Bob's) input is $0$ ($0$). Suppose the Bell value of the CHSH inequality is $J_{CHSH}$, the relation is
\begin{equation}
	J_{CHSH} = -J_{CH}/2.
\end{equation}

The prepared quantum state is in the form of
\begin{equation}\label{eq:Eberhardstate}
  \ket{\psi} = \frac{1}{\sqrt{1+r^2}}(\ket{HV} + r\ket{VH}).
\end{equation}
The measurement of Alice and Bob are projective measurement in the following basis
\begin{equation}
	\{\cos{\alpha}\ket{H}+\sin{\alpha}\ket{V}, -\sin{\alpha}\ket{H} + \cos{\alpha}\ket{V}\}.
\end{equation}
With given detection efficiency, the optimal state and measurement can be numerically found.

Apart from the detection efficiency, other  experimental imperfections such as imperfect photon source, background  noise, misalignment error should be taken into account in order to simulate the experiment.

\subsection{Imperfect input state}
To achieve an optimal violation of the CH inequality, we prepare quantum state in the form of Eq.~\eqref{eq:Eberhardstate}.
In practice, the experimentally prepared input state may be imperfect, deviating from the desired one. One way is to model the imperfectly prepared state by
\begin{equation}\label{Eq:visibility}
  \rho = \frac{1}{1+r^2}\left(
                          \begin{array}{cccc}
                            0 & 0 & 0 & 0 \\
                            0 & 1 & Vr & 0 \\
                            0 & Vr & r^2 & 0 \\
                            0 & 0 & 0 & 0 \\
                          \end{array}
                        \right),
\end{equation}
where $V$ can be interpreted as the visibility along a certain measurement direction. Thus a smaller $V$ represents a larger imperfection. In experiment, we can also perform  a state tomography to directly get an estimation of the prepared state.

\subsection{SPDC source}
The photon pair distribution of an SPDC source follows a Poisson distribution,
\begin{equation}\label{}
  P(n) = \frac{\mu^n}{n!}e^{-\mu}.
\end{equation}
or a thermal distribution,
\begin{equation}\label{}
  P(n) = \frac{\mu^n}{(1+\mu)^{n+1}}e^{-\mu},
\end{equation}
where $\mu$ is the mean photon number.
In this case, the source may emit vacuum with probability $P(0)$, one pair with probability $P(1)$, two pairs with probability $P(2)$, and so on. When considering no background, the  vacuum part contributes $0$ to the Bell value and the rest part contributes positive Bell value.

In the following, we will evaluate the Bell value with multiple photon pairs.  As the probability of multiple photon pairs is very small, we only calculate the averaged Bell value ${J}_2$ with two photon pairs as input. We consider threshold detector in our simulation.

In general, there are two detectors for each measurement. For single photon pairs, there are nine exclusive events as shown in Table~\ref{Table:twopair}.
\begin{table}[hbt]
\centering
\caption{Possible events for single photon pair. }
\begin{tabular}{c|ccccccccc}
  \hline
  parties&1 &2 & 3 & 4 &5&6&7&8&9\\
  \hline
  Alice& 0 & 0 & 0 & 1 & 1 & 1 & u & u & u\\
  Bob& 0 & 1 & u & 0 & 1 & u & 0 & 1 & u\\
  \hline
\end{tabular}\label{Table:twopair}
\end{table}
For two photon pairs, there are $9\times9=81$ exclusive events.
However, when threshold detectors are used, there are only 4 different distinguishable events for Alice (Bob), including, 0, 1, u, and double click (detecting 0 and 1 at the same time). As the output of the original CH inequality has only three distinct outputs for each party, we thus need to assign an output (0, 1, or u) for the events of double clicks.

Generally,  when a double click happens, we randomly assign the output value to be 0, 1, or u, with probability $q_0$,  $q_1$, and $q_u$, respectively. For example, when the 00 and 01 events happens for the two pairs, Alice will detect oo and thus also output o, while Bob will detect 01 and output 0 or 1 with additional probability $q_0$ or $q_1$, respectively. When $q_0=1$ it corresponds to the case that double click events are all assigned to output of 0 which corresponds to the case that only one detector is used for each party.

After the random assignment, the output contains nine exclusive events. Note that, by saying double click, we actually mean the events of detecting 0 and 1 at the same time. For the other cases, for instance, when detecting 0 and u or 0 and 0 at the same time, we simply assign the output to be 0 deterministically. This is because that we cannot distinguish between the events of 0u, u0, and 00 for each party in experiment.

It is worth mentioning that the random assignment must be performed instantly after a measurement outcome is observed, such that Alice and Bob cannot communicate. That is, the events of random assignments of Alice and Bob must be space-likely separated. In experiment, such requirement can be satisfied by considering that random assignment is delay-processed in data postprocessing. In this case, the random assignment strategies of Alice and Bob should be fixed before experiment. This requirement can be fulfilled by introducing a trusted referee to perform the data postprocessing. He will honestly perform the random assignment of the data from Alice and Bob with a predetermined strategy. 

Suppose the underlying probability distribution of the nine events with single photon pair conditioned on specific input setting $x$ and $y$ are denoted by $p_i(x, y)$ for experimental trial $i$,  we can calculate the probability $p_1^{n=2}(x, y)$ of obtaining 00 with two pairs of photons after the assignment by
\begin{equation}\label{eq:twopairab}
  p_1^{n=2}(x, y) = \sum_{i,j}\beta_{ij}p_i(x, y)p_j(x, y),
\end{equation}
where the coefficients $\beta_{ij}$ are given by
\begin{equation}\label{}
  \beta_{ij} =
  \left(
    \begin{array}{ccccccccc}
      1 & q_0^a & 1 & q_0^b & q_0^aq_0^b & q_0^a & 1 & q_0^b & 1\\
      q_0^a & 0 & 0 & q_0^aq_0^b & 0 & 0& q_0^b& 0& 0\\
      1 & 0& 0& q_0^a& 0& 0& 1& 0&0\\
      q_0^b & q_0^aq_0^b& q_0^a& 0& 0& 0& 0& 0&0\\
      q_0^aq_0^b & 0& 0& 0& 0& 0& 0& 0&0\\
      q_0^a & 0& 0& 0& 0& 0& 0& 0&0\\
      1 & q_0^b& 1& 0& 0& 0& 0& 0&0\\
      q_0^b & 0& 0& 0& 0& 0& 0& 0&0\\
      1 & 0& 0& 0& 0& 0& 0& 0&0\\
    \end{array}
  \right),
\end{equation}
where $q_0^a$ and $q_0^b$ refer to the random assignment probability of Alice and Bob, respectively.

For single party detections, there are three types of events: 0, 1, and u.  When there are two pairs, we can similarly assign double click events. For instance, for single detection probability $p_1^A(x)$, where $1$ also stands for event o, we have the corresponded one for two pairs,
\begin{equation}\label{eq:twopaira}
  p_1^{A,n=2}(x) = \sum_{i,j} \beta_{ij}' p_i^A(x)p_j^A(x),
\end{equation}
with coefficient defined by
  \begin{equation}\label{}
  \beta_{ij}' =
  \left(
    \begin{array}{ccc}
      1 & q_0^a & 1 \\
      q_0^a & 0 & 0 \\
      1 & 0& 0\\
    \end{array}
  \right)
\end{equation}

Thus, we can calculate the normalized contribution from two pairs, denoted by $\bar{J}_{n=2}$ according to Eq.~\eqref{eq:twopairab} and Eq.~\eqref{eq:twopaira}. Then a weighted Bell value is given by
\begin{equation}\label{}
  J \approx P(1){J}_{n=1} + P(2){J}_{n=2}.
\end{equation}

\subsection{Misalignment error}
Due to imperfection in detection, there may also be misalignment error which outputs value opposite to the real one. Suppose the rate of misalignment error is $p_M$, then each party will output 0 (1) when its actual value is 1 (0) with probability $p_M$. Denote the events by the index in Table.~\ref{Table:twopair}, then the coincidence probability after misalignment is

\begin{equation}\label{*}
  p_i^{M}(x, y) = \sum_{i}\gamma_{ij}p_j(x, y), \forall i
\end{equation}
where the coefficients $\gamma_{ij}$ is given by
\begin{widetext}
	\begin{equation*}\label{}
\begin{aligned}
  &\gamma_{ij} =\\
  &\left(
    \begin{array}{ccccccccc}
       (1-p_M)^2 &  (1-p_M)p_M & 0 &  (1-p_M)p_M & p_M^2 & 0 & 0 & 0 & 0\\
       (1-p_M)p_M &  (1-p_M)^2 & 0 & p_M^2 & (1-p_M)p_M & 0& 0& 0& 0\\
      0 & 0& (1-p_M)& 0& 0& p_M& 0& 0&0\\
      (1-p_M)p_M & p_M^2& 0& (1-p_M)^2& (1-p_M)p_M& 0& 0& 0&0\\
      p_M^2 & (1-p_M)p_M& 0& (1-p_M)p_M& (1-p_M)^2& 0& 0& 0&0\\
      0 & 0& p_M& 0& 0& (1-p_M)& 0& 0&0\\
      0 & 0& 0& 0& 0& 0& (1-p_M)& p_M&0\\
      0 & 0& 0& 0& 0& 0& p_M& (1-p_M)&0\\
      0 & 0& 0& 0& 0& 0& 0& 0&1\\
    \end{array}
  \right),
\end{aligned}
\end{equation*}
\end{widetext}

For single detections, such as $p_0^A(x)$, we similarly have
 \begin{equation}\label{}
  p_{0}^{A,M}(x) = (1-p_M)p_{0}^A(x) + p_Mp_{1}^A(x).
\end{equation}

\subsection{Dark count}
In this part, we consider how background affects the experiment. Suppose the background is described by $p_B$, which denotes the probability of detecting a dark count for each detector. For instance, when the number of emitted pulses  and  dark counts are $n$  and $B$ for each detector, then $p_B = B/n$.

Due to the existence of dark counts, the Bell value will be affected when the input is vacuum. For coincidence detection $p_{00}(x,y)$, the probability contributed from dark counts is $p^{B,n=0}_{00}(x,y) = p_B^2$; for single detection $p_0^A(x)$, the probability contributed from dark counts is $p^{B,n=0}_{0}(x) = p_B$.

For the inputs with nonzero photon pairs, the situation becomes more complicated. For instance, when the source emit one photon pair, the  probability contributed from dark counts of coincidence detection $p_{00}(x,y)$ is
\begin{equation}\label{}
\begin{aligned}
  p^{B,n=1}_{00}(x,y) &= p_{00}(x, y)\eta_a(1-\eta_b)p_B + p_{00}(x, y)(1-\eta_a)\eta_bp_B\\
  & + p_{00}(x, y)(1-\eta_a)(1-\eta_b)p_B^2 + (1-p_{00}(x, y))p_B^2,
\end{aligned}
\end{equation}
where the four terms  correspond to four cases of the detection of the photon pair: Bob detects vacuum due to loss, Alice detect vacuum due to loss, Alice and Bob both detect vacuum due to loss, Alice and Bob detect other events. Here, we only consider that Alice and Bob each use one detector. The result is only slightly different when they each use two detectors.

Denote the normalized contribution from dark count by ${J}_B$, then the total Bell value $J$ is given by
\begin{equation}\label{}
  J \approx {J}_B+ P(1){J}_{n=1} + P(2){J}_{n=2}.
\end{equation}

In our simulation, we first generate the probability distribution by using the quantum state from tomography. Then, we consider misalignment error and multi-photon states. Finally, we take into account the contribution from background noise.

\begin{figure}[tbh]
\centering
     \resizebox{9cm}{!}{\includegraphics{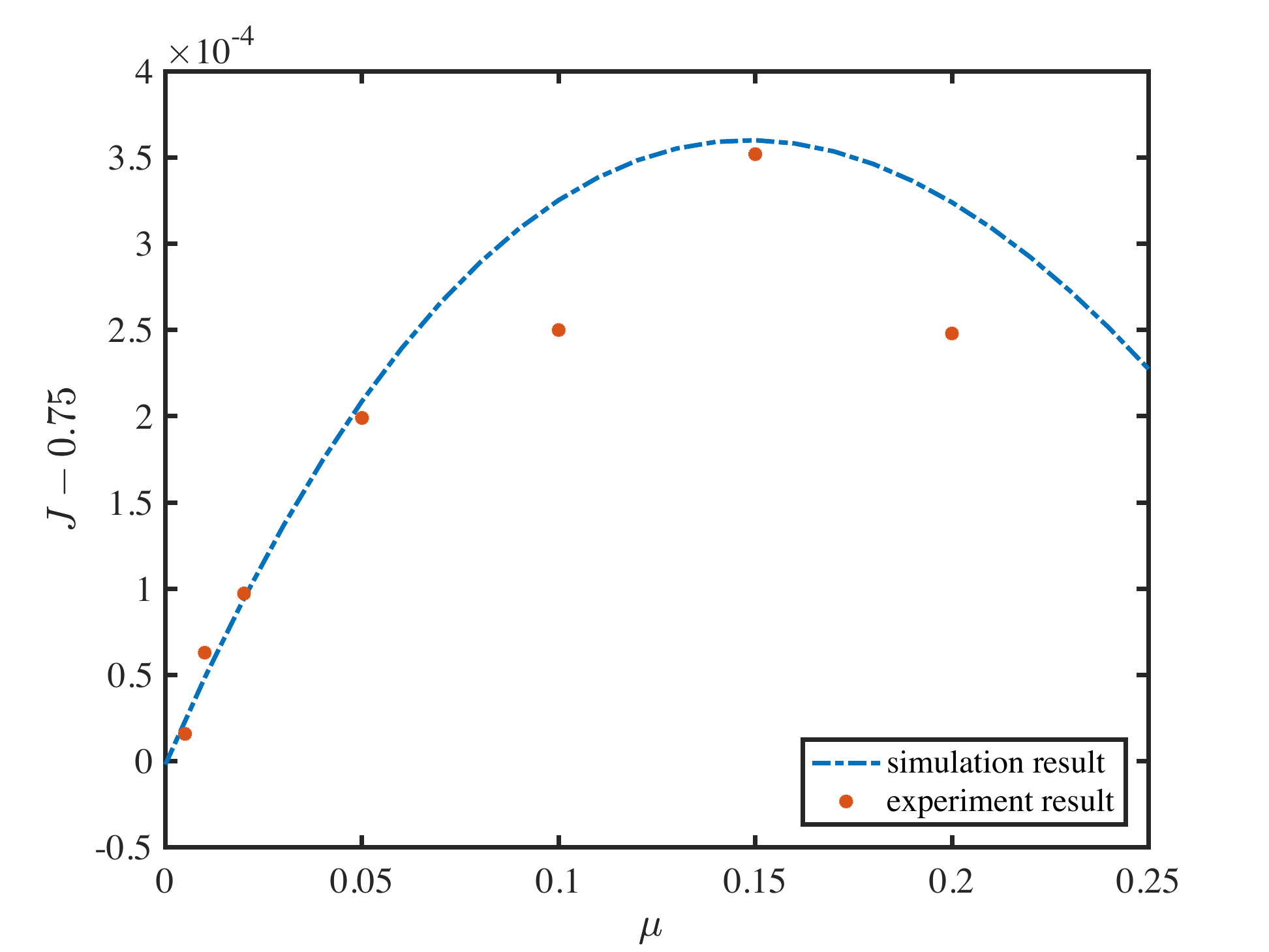}}
     \caption{Simulation result. The relation between Bell inequality violation and mean photon number. In our simulation, we set $p_B = 2\times10^{-6}$ and $p_M = 0.002$.}
\label{Fig:randombbits}
\end{figure}

\section{Randomness extraction}
Toeplitz extractor  is used for randomness extraction from raw data \cite{Impagliazzo:Leftover:1989, frauchiger2013true}. In total, we have $4\times10^{10}$ experimental trials. 2 bits are used to record the detection results in each trial. For instance, $a=0,b=1$ represent that Alice does not detect a photon while Bob detects one photon, which is noted as $ab=01$ in the text. As a result, $8\times10^{10}$ bits of raw data are collected.  Multiplying a Toeplitz matrix with dimensions $m \times n =(4.56\times10^7) \times (8\times10^{10})$ to the raw data vector with a dimension $n=8\times10^{10}$ yields a vector with a dimension $m=4.56\times10^7$.

A $m\times n$ Toeplitz matrix takes the from:

\begin{equation}\label{}
  T_{m\times n} =
  \left(
    \begin{array}{ccccc}
    a_0     & a_{-1}  & \cdots & a_{-(n-2)}   & a_{-(n-1)} \\
    a_1     & a_0     & \ddots &              & a_{-(n-1)+1} \\
    a_2     & a_1     & \ddots & \ddots       & \vdots     \\
    \vdots  & \vdots  &        & \ddots       & a_{-(n-1)+(m-2)} \\
    a_{m-1} & a_{m-2} & \cdots & a_{-n+(m-1)} & a_{-(n-1)+(m-1)} \\
    \end{array}
  \right)
\end{equation}
with raw data vector:
\begin{equation}\label{}
  V_{n} =
  \left(
    \begin{array}{c}
    v_0 \\
    v_1 \\
    v_2 \\
    \vdots  \\
    v_{n-1} \\
    \end{array}
  \right)
\end{equation}
The result is:
\begin{equation}\label{}
  R_{m} =
  \left(
    \begin{array}{c}
    r_0 \\
    r_1 \\
    r_2 \\
    \vdots  \\
    r_{m-1} \\
    \end{array}
  \right)
\end{equation}
The result random number is obtained by a multiplication between a Toeplitz matrix and a vector:
\begin{equation}
	R_{m}=T_{m\times n}\times V_{n}
\end{equation}

An fast Fourier transform (FFT) can be applied to speed up the multiplication:
\begin{equation}
	T_{m\times n}\times V_n = IFFT( FFT(T_{m+n-1}) \cdot FFT(V_m) )
\end{equation}

Here FFT is the fast Fourier transform on the vector, $T_{m+n-1}$ is the elements $(a_{-(n-1)}, ... , a_{-1}, a_0, a_1, ..., a_{m-1})$ in the Toeplitz matrix. IFFT is the inverse fast Fourier transform of the product of the vectors. In calculation, the vector dimension should be expand to $m+n-1$ by adding zeros at the end.

In our experiment, we divide the matrix into $k=n/l$ blocks each with dimension $m\times l$:

\begin{equation}\label{}
  T_{m\times n} =
  \left(
    \begin{array}{cccc}
    T_{m\times l}^0 & T_{m\times l}^1 & \cdots & T_{m\times l}^{k-1}
    \end{array}
  \right)
\end{equation}
and the block $T_{m\times l}^{i}$ is:
\begin{equation}\label{}
  T_{m\times l}^{i} =
  \left(
    \begin{array}{cccc}
    a_{-i\cdot l}     & a_{-(i\cdot l+1)} & \cdots   & a_{-(i\cdot l+l-1)}     \\
    a_{-i\cdot l+1}   & a_{-i\cdot l}     & \ddots   & a_{-(i\cdot l+l-1)+1}   \\
    \vdots            & \vdots            &          & \vdots                  \\
    a_{-i\cdot l+m-1} & a_{-i\cdot l+m-2} & \cdots   & a_{-(i\cdot l+l-1)+m-1} \\
    \end{array}
  \right)
\end{equation}

Similarly, the vector is divided into k blocks:
\begin{equation}\label{}
  V_{n} =
  \left(
    \begin{array}{c}
    V_l^0 \\
    V_l^1 \\
    \vdots \\
    V_l^{k-1}
    \end{array}
  \right)
\end{equation}
with each block:
\begin{equation}\label{}
  V_{l}^i =
  \left(
    \begin{array}{c}
    v_{i\cdot l} \\
    v_{i\cdot l+1} \\
    \vdots \\
    v_{i\cdot l+l-1} \\
    \end{array}
  \right)
\end{equation}

For each block, FFT speed up can be applied, and a series of blocks of result is obtained:
\begin{equation}\label{}
  R_{m}' =
  \left(
    \begin{array}{cccc}
    R_l^0 & R_l^1 & \cdots & R_l^{k-1}
    \end{array}
  \right)
\end{equation}
where $R_l^i=T^i_{m\times l}\cdot V^i_l$, with each block:
\begin{equation}\label{}
  R_{l}^i =
  \left(
    \begin{array}{cc}
    r_0^i\\
    r_1^i\\
    \vdots\\
    r_{m-1}^i\\
    \end{array}
  \right)
\end{equation}
Finally, each row of the result block is added to give the result:
\begin{equation}\label{}
  R_{m} =
  \left(
    \begin{array}{cc}
    \Sigma_i r_0^i\\
    \Sigma_i r_1^i\\
    \vdots\\
    \Sigma_i r_{m-1}^i\\
    \end{array}
  \right)
\end{equation}
The blocked algorithm is slower than the full FFT algorithm, but it saves memory. We perform the extraction calculation on a personal computer with 16 Gbytes memory by dividing the original data into 800 blocks, which takes 17 hrs including data loading and computation.

\section{Statistical analysis}

We obtain $4.56\times10^7$ random bits. We use 45 M bits data with section length set to 450 K bits for NIST statistical test \cite{NIST_Tests}. As shown in Tab.~\ref{tab:nisttest}, the random bits successfully pass the NIST tests.

\begin{table}[htb]
\centering
  \caption{ Results of the NIST test suite using 45 Mbit of data (100 sequences of 450 Kbit) with the generated random numbers. }
\begin{tabular}{c|ccc}
\hline
Statistical tests & P value & Proportion & Result\\
\hline
Frequency				& 0.63712 & 1.000 & Success \\
BlockFrequency			& 0.22482 & 0.980 & Success \\
CumulativeSums			& 0.37040 & 1.000 & Success \\
Runs					& 0.33454 & 1.000 & Success \\
LongestRun				& 0.61631 & 1.000 & Success \\
Rank					& 0.30413 & 0.990 & Success \\
FFT						& 0.77919 & 1.000 & Success \\
NonOverlappingTemplate	& 0.71975 & 0.991 & Success \\
OverlappingTemplate		& 0.09320 & 1.000 & Success \\
Universal				& 0.36692 & 0.970 & Success \\
ApproximateEntropy		& 0.51412 & 1.000 & Success \\
RandomExcursions 		& 0.38200 & 0.995 & Success \\
RandomExcursionsVariant	& 0.41480 & 0.995 & Success \\
Serial 					& 0.42510 & 0.995 & Success \\
LinearComplexity		& 0.91141 & 0.990 & Success \\
\hline
\end{tabular}
\label{tab:nisttest}
\end{table}

\begin{figure*}[tbh]
\centering
\resizebox{10cm}{!}{\includegraphics{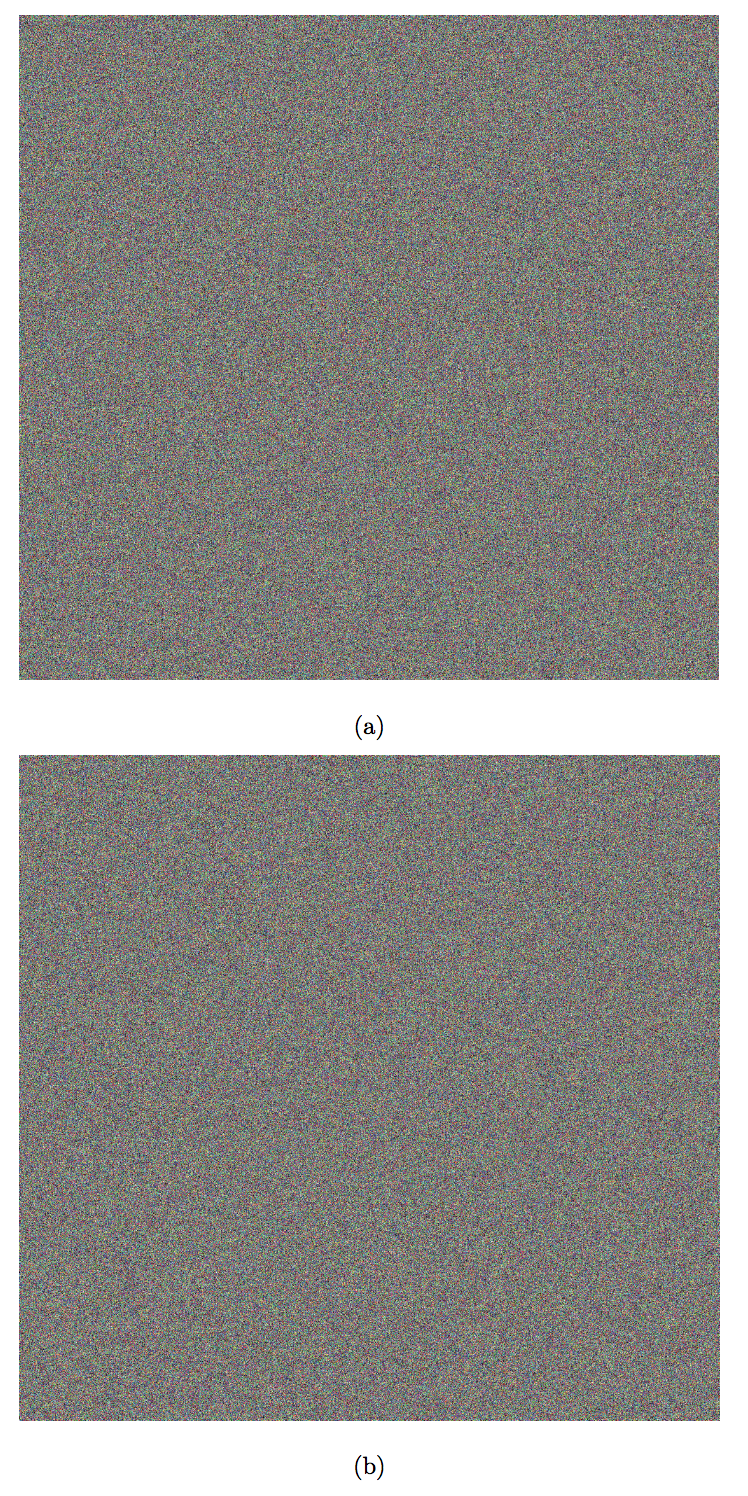}}
\caption{Then final random numbers are divided into two parts (a) and (b), each figure contains $1000\times950$ pixels. For each pixel, 3 bytes random data are used to control the RGB color. All the random numbers generated in the experiment are used to draw the two figures.}
\label{Fig:randombbits}
\end{figure*}

\bibliography{BibDIQRNG}

\end{document}